\documentclass[11pt]{article}
\usepackage{epsfig} 
\usepackage{amssymb}
\usepackage{graphics}
\newcommand{\sect}[1]{ \section{#1} \setcounter{equation}{0} } 

\newcommand{\half}{\mbox{\small{$\frac{1}{2}$}}}

\newcommand{\Nc}{N_{\!c}}
\newcommand{\Nf}{N_{\!f}}

\newcommand{\NA}{N_{\!A}}
\newcommand{\bare}{\mbox{\footnotesize{o}}}

\newcommand{\Dslash}{D \! \! \! \! /}
\newcommand{\lin}{\mbox{lin}}
\newcommand{\linss}{\mbox{\scriptsize{lin}}}
\newcommand{\MSbar}{\overline{\mbox{MS}}}
\newcommand{\MSbars}{\overline{\mbox{\footnotesize{MS}}}}
\newcommand{\MSbarss}{\overline{\mbox{\scriptsize{MS}}}}
\newcommand{\mMOM}{\mbox{mMOM}}

\newcommand{\mMOMss}{\mbox{\scriptsize{mMOM}}}

\newcommand{\MOMts}{\widetilde{\mbox{\footnotesize{MOM}}}}

\begin{document}

\title{Four loop renormalization of QCD in the Curci-Ferrari gauge}

\author{J.A. Gracey, \\ Theoretical Physics Division, \\ 
Department of Mathematical Sciences, \\ University of Liverpool, \\ P.O. Box 
147, \\ Liverpool, \\ L69 3BX, \\ United Kingdom.} 

\date{}

\maketitle 

\vspace{5cm} 
\noindent 
{\bf Abstract.} We renormalize Quantum Chromodynamics (QCD) when gauge fixed in
the nonlinear Curci-Ferrari gauge to four loops in the modified minimal 
subtraction ($\MSbar$) scheme. We reproduce the four loop QCD $\MSbar$
$\beta$-function from the Slavnov-Taylor identity for this gauge which relates
the coupling constant renormalization to the gluon, Faddeev-Popov ghost and 
gauge parameter anomalous dimensions. This is carried out for a nonzero gauge
parameter, without having to evaluate a vertex function. The anomalous 
dimension of the BRST invariant dimension two gluon and ghost mass term is
deduced from a similar Slavnov-Taylor identity for this gauge. Consequently we
construct the renormalization group functions in the minimal momentum
subtraction scheme to four loops. As a corollary we deduce the five loop 
$\beta$-function and quark mass anomalous dimensions in the same scheme. We 
also outline the pros and cons of employing the Curci-Ferrari gauge to access 
the six loop QCD $\beta$-function in the $\MSbar$ scheme.

\vspace{-18.2cm}
\hspace{13.4cm}
{\bf LTH 1388}

\newpage 

\sect{Introduction.}

One of the problems in the early adaptation of gauge theories to particle
physics was that of how to endow vector bosons with a mass in such a way that 
the gauge principle was not violated. While the weak sector can be successfully
accommodated by massive $W$ and $Z$ bosons through the mechanism of spontaneous
symmetry breaking the properties of the gluons in the strong sector, especially
in the infrared region, remain under study. The major problems that are not 
comprehensively understood are those of colour and quark confinement as well as
chiral symmetry breaking. For the first item lattice field theory has proved 
instrumental in revealing interesting properties of the gluon propagator. For 
instance, several decades ago the Landau gauge gluon propagator was shown to 
freeze at a nonzero value in the infrared limit, \cite{1,2,3,4,5,6,7,8}. To be
clear this does not imply the gluon has a mass primarily since the propagator 
does not have a simple pole at any momentum value. Instead it indicates that 
there is some fundamental scale associated with low momenta. This behaviour has
been examined from other directions. Indeed Dyson-Schwinger equation studies, 
\cite{9,10,11,12,13,14,15}, have incorporated a dynamical gluon mass which 
reproduces the lattice analyses. Another analytic approach based on 
perturbative quantum field theory uses a Landau gauge Lagrangian where the 
gluon has a mass term. The inclusion of such a mass for this gauge was 
developed by Curci and Ferrari in \cite{16}. There a Becchi-Rouet-Stora-Tyutin 
(BRST) invariant Lagrangian was constructed with a BRST invariant gluon and 
ghost mass term. When the gauge parameter of the model is set to zero the 
Landau gauge fixed Lagrangian is produced. Although BRST symmetry is clearly 
not gauge symmetry the Curci-Ferrari model does have several beneficial 
properties. Using the Landau gauge setup in \cite{17,18,19,20} the one and two
loop gluon and ghost propagators were constructed and compared with the lattice
results. With the running coupling and gluon mass as the only two input 
parameters a very accurate fit of the lattice data was found over all momentum 
scales. In some ways this is an interesting observation for a perturbative 
computation. 

While a mass term may seem to contradict the breaking of gauge symmetry its 
presence appears to ameliorate the passage to the infrared region. Indeed a 
credible origin of the gluon mass term was established in a series of articles 
\cite{17,22,23,24}. It was argued in \cite{24} that the gluon mass had an 
interpretation as an additional gauge parameter that measures or accounts for 
Gribov copies that were discussed orginally in \cite{25}. Such copies arise due
to the inability to {\em globally} fix a covariant gauge uniquely. In a
Yang-Mills context such copy problems arise in the infrared limit, \cite{25}. 
While Gribov constructed a Lagrangian that accommodated copies the restriction 
of the gluon field to a subspace of configuration space introduced a mass 
scale. This satisfied a gap equation meaning that the Gribov mass was not an 
independent parameter of the theory. Its anomalous dimension is a linear 
combination of the ghost and gluon dimensions in the Landau gauge. The 
resulting Landau gauge gluon propagator, although being nonfundamental, tended
to zero in the infrared limit, \cite{25}, different from current lattice 
analyses. However, the relevant point is that a mass term for the gluon in the 
Landau gauge may in fact not be as problematic as initially thought. Indeed one
can include a gauge {\em invariant} dimension two gluon mass operator as shown 
in \cite{9,26,27} by slackening the requirement of locality.

Providing this context for the Curci-Ferrari model justifies its treatment at 
least as a laboratory to explore gauge symmetry properties. However it has
features that are applicable to problems requiring high loop order results. For
instance, the Lagrangian of \cite{16} has a different ghost sector from that of
the Faddeev-Popov construction for the linear gauge since it contains a quartic
ghost interaction which does not invalidate renormalizability. Also the gauge 
parameter, $\alpha$, of \cite{16} is not synomymous with that of the 
canonically employed linear covariant gauge. It is important to make this 
distinction at the outset. It lies in the fact that the gauge fixing functional
of the canonical gauge is linear whereas that of \cite{16} is nonlinear in the 
fields and $\alpha$ dependent. Therefore when $\alpha$~$=$~$1$ in the 
Curci-Ferrari model it is not the Feynman gauge of the linear covariant gauge.
Another difference between the linear and nonlinear gauge is that the 
Curci-Ferrari gauge\footnote{The gauge was also studied in \cite{28} and is
sometimes termed the Curci-Ferrari-Delbourgo-Jarvis gauge.} has different 
Slavnov-Taylor identities which have implications for the relations between the
renormalization constants. For instance it was proved in \cite{16,29,30} that 
the BRST invariant dimension two gluon mass operator does not have an 
independent renormalization constant. Instead it is related to the gluon, ghost
and gauge parameter renormalization constants. Unlike the linear covariant 
gauge the Curci-Ferrari gauge parameter renormalization constant is not unity 
as is evident from explicit computations, \cite{31}. The formal construction of 
\cite{29} was motivated by the fact that the relation was revealed by a three 
loop computation in \cite{32}. Indeed the more general result of \cite{29,30} 
incorporates similar relations observed in other three loop computations for 
the Landau gauge in \cite{31} although a proof had been noted earlier in 
\cite{33} and later reproduced in \cite{34}. In the Landau gauge case the mass 
operator dimension is the sum of the ghost and gluon dimensions. In other words
its renormalization is the same as that of the Gribov mass \cite{35,36}. One 
unusual aspect of this property is that the Gribov mass parameter is associated
with a dimension four {\em nonlocal} operator in the Gribov Lagrangian, 
\cite{25}, whereas the BRST invariant mass operator of \cite{16} is necessarily
dimension two.

Another salient feature of the Curci-Ferrari model also concerns a 
Slavnov-Taylor identity which is the one that generalizes Taylor's Landau gauge
theorem, \cite{37}. That result states that the ghost-gluon vertex is finite in
the Landau gauge to all orders in perturbation theory. This implies that the
coupling renormalization constant can be determined solely from the ghost and
gluon field renormalization constants in the Landau gauge. By contrast the
equivalent Slavnov-Taylor identity in the Curci-Ferrari gauge, \cite{29,30}, 
expresses the coupling renormalization constant in terms of not only the ghost 
and gluon renormalization constants but additionally that of the gauge 
parameter which is not unity in this gauge. On the one hand this means that the 
$\beta$-function can be deduced without evaluating a vertex function in much
the same way the Taylor theorem implies in the Landau gauge. More significantly
it means that one can deduce the $\beta$-function when the Curci-Ferrari gauge
parameter $\alpha$ is unity. For this value the gluon propagator simplifies to 
a standard scalar propagator. These insights could have important implications 
for any attempt to compute the $\beta$-function of Quantum Chromodynamics (QCD)
beyond the current five loop result of \cite{38,39,40,41}. In other words the 
excessive integration by parts associated with the dipole part of the gluon 
propagator is immediately circumvented at the outset. While this would
substantually reduce the amount of work required to evaluate a high order 
$\beta$-function it does not of course decrease the number of Feynman graphs to
be computed for the ghost and gluon $2$-point functions. It does, however, 
exclude having to evaluate a vertex function. Although there will be more 
$2$-point graphs in the Curci-Ferrari gauge due to the presence of the quartic 
ghost vertex, compared to the linear covariant gauge, the interaction itself is 
akin to that of scalar $\phi^4$ theory and does not have any momenta.

In explaining this background to the nonlinear Curci-Ferrari gauge it is the
aim of the article to renormalize QCD with this gauge fixing to four loops. 
This will extend the lower loop results of \cite{31}. As a corollary we will 
demonstrate the benefit of this gauge choice in reproducing the four loop QCD 
$\beta$-function in the $\MSbar$ scheme, \cite{42,43}. To achieve this new 
order we will employ the {\sc Forcer} algorithm, \cite{44,45}, written in the 
highly efficient symbolic manipulation language {\sc Form}, \cite{46,47}. One 
consequence of this approach is that we will also carry out the renormalization
in the minimal momentum subtraction ($\mMOM$) scheme of \cite{48}. This scheme 
was first introduced for the canonical linear covariant gauge as a scheme which
seeks to preserve the core feature of Taylor's theorem for the ghost-gluon 
vertex for linear covariant gauges, \cite{37,49} but beyond that of the Landau 
gauge. It will turn out that due to the Slavnov-Taylor identity of \cite{29,30} 
for the Curci-Ferrari gauge the $\mMOM$ version of the QCD $\beta$-function has
some natural structural properties for a nonzero gauge parameter. As the five 
loop $\MSbar$ QCD $\beta$-function is known, \cite{38,39,40,41}, and is 
independent of the gauge parameter, \cite{50}, we will determine the five loop 
$\mMOM$ $\beta$-function in the Curci-Ferrari gauge by exploiting properties of
the renormalization group. A similar analysis will be carried out for the quark
mass anomalous dimension which is equally independent of the gauge parameter in
the $\MSbar$ scheme, \cite{50}. In addition providing the anomalous dimensions 
of the fields and gauge parameter to four loops in the Curci-Ferrari gauge in 
both the $\MSbar$ and $\mMOM$ schemes will be important in refining the 
critical point analysis of the renormalization group flow recently carried out 
in \cite{51}.

The article is structured as follows. Section $2$ is devoted to the properties
of the Curci-Ferrari model and a general outline of its renormalization
properties including the strategy we will use to find the four loop
renormalization group functions. These are provided in Sections $3$ and $4$ in
the $\MSbar$ and $\mMOM$ schemes respectively with the five loop $\mMOM$
$\beta$-function and quark mass operator anomalous dimension being recorded in 
the latter section. The potential usefulness of the Curci-Ferrari gauge in 
going beyond five loops is discussed in Section $5$ where the pros and cons of
two main strategies are outlined. Concluding remarks are provided in Section 
$6$. Two Appendices follow with Appendix A giving the $\MSbar$ field anomalous 
dimensions for an arbitrary colour group and gauge parameter in the 
Curci-Ferrari gauge. Appendix B contains the five loop $\mMOM$ $\beta$-function
for the same general case.

\sect{Background.}

First we introduce the QCD Lagrangian gauge fixed in the Curci-Ferrari gauge
\cite{16}. In terms of bare quantities it is
\begin{eqnarray}
L &=& -~ \frac{1}{4} G^a_{\bare \,\mu\nu} G_{\bare}^{a \, \mu\nu} ~+~ 
i \bar{\psi}^i_{\bare} \Dslash_{\bare} \psi_{\bare}^i ~-~
\frac{1}{2\alpha_{\bare}} \left( \partial_\mu A^{a\,\mu}_{\bare} \right)^2 ~+~
\bar{c}_{\bare}^a \left( \partial_\mu D^\mu_{\bare} c_{\bare} \right)^a 
\nonumber \\
&& -~ \frac{g_{\bare}}{2} f^{abc} \bar{c}_{\bare}^a c_{\bare}^b 
\,\, \partial^\mu \! A^c_{\mu \, \bare} ~+~ 
\frac{1}{8} \alpha_{\bare} g_{\bare}^2 f^{abcd} 
\bar{c}_{\bare}^a c_{\bare}^b \bar{c}_{\bare}^c c_{\bare}^d 
\label{lagcf}
\end{eqnarray}
where $A^a_\mu$, $c^a$ and $\psi^i$ are the gluon, Faddeev-Popov and quark
fields respectively. Also $g$ is the gauge coupling, $\alpha$ is the gauge 
parameter, $G^a_{\mu\nu}$ is the gluon field strength, $D_\mu$ is the covariant
derivative and the index ranges are $1$~$\leq$~$i$~$\leq$~$\Nf$, 
$1$~$\leq$~$a$~$\leq$~$\NA$ and $1$~$\leq$~$I$~$\leq$~$\Nc$ where $\Nf$ is the 
number of quark flavours and $\NA$ and $\Nc$ and the respective dimensions of 
the adjoint and fundamental representations. The colour group generators are 
denoted by $T^a_{IJ}$. The subscript ${}_{\bare}$ on an object indicates it is 
bare or a function of bare quantities like a field or $g$. The Lagrangian 
(\ref{lagcf}) was placed on a firmer foundation in \cite{52,53,54} where the 
gauge fixing sector was shown to be equivalent to the application of an 
anti-BRST transformation, \cite{29,30,55}, to a dimension two operator followed
by a BRST one. The key differences between (\ref{lagcf}) and its linear 
covariant gauge fixed counterpart are the presence of a quartic ghost 
interaction and a different ghost-gluon interaction. Neither term affects the 
renormalizability of the Lagrangian, \cite{16}. However (\ref{lagcf}) does have
a connection with the linear covariant gauge in that when the parameter 
$\alpha$ is set to zero one obtains the Landau gauge fixed Lagrangian. While 
the ghost-gluon interaction of the conventional Faddeev-Popov construction is 
contained within the final term of the first line the full interaction has the 
additional contribution from the first term of the second line. That term 
clearly vanishes under the Landau gauge condition 
$\partial^\mu A^a_\mu$~$=$~$0$. We note that in the original article \cite{16} 
a more general Lagrangian appeared which depended on an additional parameter 
being denoted by $\beta$ there. It was not an extra gauge parameter as such but
instead acted as an interpolating entity between (\ref{lagcf}) and the linear 
gauge. One interesting feature of this particular Curci-Ferrari gauge fixing is
that it admits a dimension two BRST invariant mass operator ${\cal O}$, 
\cite{16}, where 
\begin{equation}
{\cal O} ~=~ \half A^a_\mu A^{a \, \mu} ~-~ \alpha \bar{c}^a c^a ~.
\end{equation}
Adding such a term to (\ref{lagcf}) with a mass parameter $m$ produces what is
known as the Curci-Ferrari {\em model} with the consequence that the ghost 
becomes massive with mass $\alpha m^2$. If the gluon propagator is decomposed 
into transverse and longitudinal modes, the former has a conventional massive
propagator with mass $m^2$ while the latter has a mass $\alpha m^2$ which 
necessarily matches that of the ghost. More concretely the respective
propagators are
\begin{eqnarray}
\langle A^a_\mu(p) A^b_\nu(-p) \rangle &=& -~ \delta^{ab} \left[
\frac{P_{\mu\nu}(p)}{(p^2+m^2)} ~+~
\frac{L_{\mu\nu}(p)}{(p^2+\alpha m^2)} \right] \nonumber \\
\langle c^a(p) \bar{c}^b(-p) \rangle &=& \frac{\delta^{ab}}{(p^2+\alpha m^2)} 
\label{cfpropsm}
\end{eqnarray}
where
\begin{equation}
P_{\mu\nu}(p) ~=~ \eta_{\mu\nu} ~-~ \frac{p_\mu p_\nu}{p^2} ~~~,~~~
L_{\mu\nu}(p) ~=~ \frac{p_\mu p_\nu}{p^2} ~.
\end{equation}
From the point of view of establishing the four loop renormalization group 
functions of (\ref{lagcf}) we will proceed with the massless formulation since 
that version can be treated with the {\sc Forcer} package, \cite{44,45}.

The bare Lagrangian is converted to one involving renormalized quantities via 
the rescalings
\begin{equation}
A^{a \, \mu}_{\bare} ~=~ \sqrt{Z_A} \, A^{a \, \mu} ~~,~~
c^a_{\bare} ~=~ \sqrt{Z_c} \, c^a ~~,~~
\psi_{\bare} ~=~ \sqrt{Z_\psi} \psi ~~,~~
g_{\bare} ~=~ \mu^\epsilon Z_g \, g ~~,~~
\alpha_{\bare} ~=~ Z^{-1}_\alpha Z_A \, \alpha
\label{zdef}
\end{equation}
which introduce the respective renormalization constants. Once these have been
determined in a specific renormalization scheme the renormalization group
functions are found from
\begin{equation}
\gamma_\phi(a,\alpha) ~=~ \left[ \beta(a,\alpha)
\frac{\partial~}{\partial a} ~+~ \alpha \gamma_\alpha(a,\alpha)
\frac{\partial~}{\partial \alpha} \right] \ln Z_\phi
\end{equation}
where $\phi$~$\in$~$\{A,c,\psi\}$ with the labels corresponding to the gluon,
ghost and quark field respectively and
\begin{equation}
\gamma_\alpha(a,\alpha) ~=~ \frac{\mu}{\alpha}
\frac{\partial \alpha}{\partial \mu} 
\end{equation}
defines the gauge parameter renormalization group function. The gauge parameter
is not included in the set $\phi$ since its renormalization is defined in terms
of two renormalization constants as given in (\ref{zdef}). This convention is 
determined from the fact that the transverse and longitudinal modes of the 
gluon $2$-point function will each have only one independent renormalization 
constant to find. These are respectively $Z_A$ and $Z_\alpha$. Consequently it 
is straightforward to see the relation between $\gamma_\alpha(a,\alpha)$, 
$Z_\alpha$ and $\gamma_A(a,\alpha)$ is
\begin{equation}
\gamma_\alpha(a,\alpha) ~=~ \left[ \beta(a,\alpha) \frac{\partial}{\partial a}
\ln Z_\alpha ~-~ \gamma_A(a,\alpha) \right] \left[ 1 ~-~ \alpha
\frac{\partial}{\partial \alpha} \ln Z_\alpha \right]^{-1} ~.
\end{equation}
This general relation contains the usual property of the linear covariant
gauge as a corollary. In that instance $Z_\alpha^{\linss}$~$=$~$1$ meaning the
gluon and gauge parameter renormalization group functions are equal and
opposite where $\lin$ will denote the linear covariant gauge throughout. By
contrast in the Curci-Ferrari gauge, \cite{31},
\begin{equation}
\gamma_A(a,\alpha) ~+~ \gamma_\alpha(a,\alpha) ~\neq~ 0 ~.
\end{equation}
In the above relations we have included $\alpha$ as an argument of the
$\beta$-function because in general it is a gauge dependent quantity. It is 
only in the $\MSbar$ scheme that it is independent of the gauge parameter,
\cite{50}.

While the three loop renormalization group functions of (\ref{lagcf}) are
available in \cite{31} an interesting property of the renormalization of 
${\cal O}$ was noted there. This was that when $\alpha$~$=$~$0$ the anomalous
dimension of ${\cal O}$ was not independent but equated to the sum of the gluon
and ghost anomalous dimensions. This relation has been established to all 
orders in the Landau gauge in \cite{33,34}. However in \cite{32} a more general
relation was recorded at three loops for $\alpha$~$\neq$~$0$. This was formally
established as a Slavnov-Taylor identity in \cite{29,30}. In particular it was 
shown that
\begin{equation}
Z_{\cal O} ~=~ \frac{Z_\alpha^2}{Z_A Z_c} ~.
\label{zop}
\end{equation}
In addition another identity was established, \cite{29,30}, which is
\begin{equation}
Z_g ~=~ \frac{Z_\alpha^2}{\sqrt{Z}_A Z_c} ~.
\label{zgccg}
\end{equation}
Both relations translate into
\begin{eqnarray}
\beta(a,\alpha) &=& a \left[ \gamma_A(a,\alpha) ~+~ 2 \gamma_c(a,\alpha) ~-~ 
4 ( \gamma_A(a,\alpha) ~+~ \gamma_\alpha(a,\alpha) ) \right] \nonumber \\
\gamma_{\cal O}(a,\alpha) &=& \gamma_A(a,\alpha) ~+~ \gamma_c(a,\alpha)
~-~ 2 ( \gamma_A(a,\alpha) ~+~ \gamma_\alpha(a,\alpha) )
\label{gammastid}
\end{eqnarray}
in terms of the respective renormalization group functions. As noted in 
\cite{29,30} (\ref{zgccg}) contains the well-known Landau gauge property that 
was established by Taylor in \cite{37,49}. One consequence of the Landau gauge 
case is that the ghost-gluon vertex is finite in that gauge. Therefore the 
$\beta$-function can be deduced from the Landau gauge ghost and gluon anomalous
dimensions. What (\ref{zgccg}) is potentially useful for, however, is in 
extracting the coupling renormalization constant at high loop order for the 
choice of $\alpha$~$=$~$1$ thereby bypassing the need to evaluate a vertex 
function. Instead only the ghost and gluon $2$-point functions would need to be
determined, \cite{29,30}, for the value $\alpha$~$=$~$1$. Finally one can 
combine the identities to express the renormalization constant of ${\cal O}$ as
\begin{equation}
Z_{\cal O} ~=~ \frac{Z_c Z_g^2}{Z_\alpha^2}
\label{zozgreln}
\end{equation}
which does not involve $Z_A$.

{\begin{figure}[ht]
\begin{center}
\includegraphics[width=12.0cm,height=5.5cm]{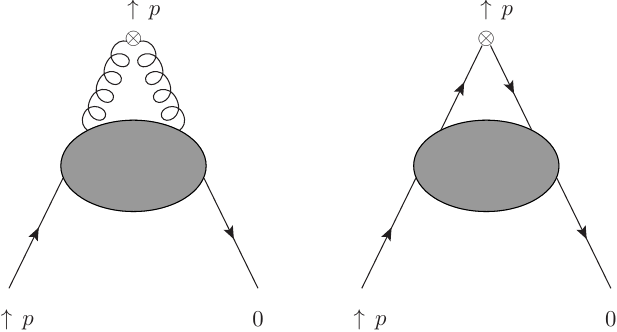}
\end{center}
\caption{Graphical structure of $G^{ab}_{{\cal O}}(p)$.}
\label{figop}
\end{figure}}

We close this section by noting we used the {\sc Forcer} package of
\cite{44,45} to carry out the task of evaluating the relevant Green's functions
to renormalize (\ref{lagcf}). The {\sc Forcer} algorithm computes the 
$\epsilon$ expansion in $d$~$=$~$4$~$-$~$2\epsilon$ dimensions of massless 
$2$-point Feynman integrals up to and including four loops. It is therefore 
ideally suited to the task and is written in the highly efficient symbolic 
manipulation language {\sc Form}, \cite{46,47}, which we used extensively. So 
the whole renormalization exercise is carried out automatically. Counterterms 
are introduced internally in summing all the graphs for a Green's function 
using the definitions (\ref{zdef}) as prescribed in \cite{56}. For the
renormalization of the BRST invariant dimension two operator the relevant 
Green's function is illustrated in Figure \ref{figop} being defined by
\begin{equation}
G^{ab}_{{\cal O}}(p) ~=~ \langle c^a(p) {\cal O}(-p) \bar{c}^b(0) \rangle ~.
\end{equation}
There are two core graphical structures as ${\cal O}$ contains two terms. The
momentum routing is chosen for several reasons. First a single external 
momentum $p$ is used to allow the use of the {\sc Forcer} package. Then the
entrance and exit points of $p$ are chosen so that one of these is at the
operator insertion itself. If the momentum flowed via the external ghost legs
then there would be potential infrared problems stemming from either operator
insertion each of whose Feynman rules is formally proportional to a unit 
operator. Moreover there is no mixing with total derivative operators. The 
particular choice of routing of the external momentum via one of the ghost legs
is not unique as the Curci-Ferrari gauge ghost-gluon vertex depends on the 
external momentum of each separate ghost fields unlike the canonical linear 
gauge. There it depends on the momentum of only one of the ghost fields. The 
operator could equally well have been inserted in a gluon $2$-point function 
corresponding to 
\begin{equation}
\hat{G}^{ab}_{{\cal O}}(p) ~=~ \langle A_\mu^a(p) {\cal O}(-p) A_\nu^b(0) 
\rangle ~.
\end{equation}
One of the reasons why we have not renormalized ${\cal O}$ using
$\hat{G}^{ab}_{{\cal O}}(p)$ is that it would introduce more gluon 
self-interaction vertices. These in turn would increase the amount of
integration needed to be carried out by {\sc Forcer} to produce the same
renormalization constant. The reason also lies in the number of Feynman graphs 
to be evaluated. We have used the {\sc Fortran} encoded {\sc Qgraf} package 
\cite{57} to generate the Feynman graphs electronically. The specific number of
graphs for the various Green's functions evaluated here is provided in Table
\ref{cfqg}. The columns labelled by a field give the number of graphs for the 
$2$-point function of that field. One reason for the enumeration of these is 
that the graph count differs from the parallel one for the linear covariant 
gauge due to the additional quartic ghost interaction of (\ref{lagcf}). The 
last two columns relate to the renormalization of ${\cal O}$ for both external 
leg choices. Aside from the extra work necessitated by the presence of extra 
gluons for $\hat{G}^{ab}_{\cal O}(p)$ compared to $G^{ab}_{\cal O}(p)$ there 
are nearly twice as many four loop graphs to be evaluated as well. Therefore 
extracting $Z_{\cal O}$ via the latter Green's function is clearly more 
efficient.

{\begin{table}[ht]
\begin{center}
\begin{tabular}{|c||r|r|r|r|r|}
\hline
\rule{0pt}{12pt}
$L$ & $\langle A^a_\mu A^b_\nu \rangle$ & 
$\langle c^a \bar{c}^b \rangle$ & 
$\langle \psi^i \bar{\psi}^j \rangle$ & $G^{ab}_{\cal O}$ & 
$\hat{G}^{ab}_{\cal O}$ \\
\hline
$1$ & $3$ & $1$ & $1$ & $3$ & $4$ \\
$2$ & $19$ & $9$ & $6$ & $44$ & $69$ \\
$3$ & $282$ & $124$ & $79$ & $949$ & $1696$ \\
$4$ & $5768$ & $2456$ & $1469$ & $25194$ & $48843$ \\
\hline
Total & $6072$ & $2590$ & $1555$ & $26190$ & $50612$ \\
\hline
\end{tabular}
\end{center}
\begin{center}
\caption{Number of graphs at each loop order $L$ for the renormalization of the
fields and ${\cal O}$ in the Curci-Ferrari gauge.}
\label{cfqg}
\end{center}
\end{table}}

\sect{$\MSbar$ scheme results.}

Having discussed the general aspects of the renormalization of (\ref{lagcf}) we
now turn to the results. First we focus on the $\MSbar$ scheme and the
renormalization of the fields and gauge parameter. The outcome of the 
{\sc Forcer} computation leads to the following anomalous dimensions
\begin{eqnarray}
\left. \gamma_A^{\MSbarss}(a,\alpha) \right|^{SU(3)} &=&
\left[
\frac{2}{3} \Nf
+ \frac{3}{2} \alpha
- \frac{13}{2}
\right] a
+ \left[
\frac{61}{6} \Nf
- \frac{531}{8}
+ \frac{9}{8} \alpha^2
+ \frac{99}{8} \alpha
\right] a^2
\nonumber \\
&&
+ \left[
\frac{8155}{36} \Nf
- 33 \zeta_3 \Nf
- \frac{29895}{32}
- \frac{215}{27} \Nf^2
+ \frac{81}{4} \zeta_3 \alpha
+ \frac{81}{64} \alpha^3
+ \frac{243}{16} \zeta_3
\right. \nonumber \\
&& \left. ~~~~
+ \frac{2727}{128} \alpha^2
+ \frac{4509}{32} \alpha
- 9 \Nf \alpha
\right] a^3
\nonumber \\
&&
+ \left[
\frac{23350603}{5184} \Nf
- \frac{10596127}{768}
- \frac{387649}{216} \zeta_3 \Nf
- \frac{170235}{128} \zeta_5 \alpha
- \frac{43033}{162} \Nf^2
\right. \nonumber \\
&& \left. ~~~~
- \frac{40905}{4} \zeta_5
- \frac{16443}{64} \zeta_4 \alpha
- \frac{10879}{36} \Nf \alpha
- \frac{8019}{32} \zeta_4
- \frac{4427}{1458} \Nf^3
- \frac{2017}{81} \zeta_3 \Nf^2
\right. \nonumber \\
&& \left. ~~~~
- \frac{1229}{216} \Nf^2 \alpha
- \frac{345}{32} \Nf \alpha^2
- \frac{135}{64} \zeta_3 \alpha^3
- \frac{45}{4} \zeta_4 \Nf \alpha
- \frac{45}{16} \zeta_3 \Nf \alpha^2
+ \frac{8}{3} \zeta_3 \Nf^3
\right. \nonumber \\
&& \left. ~~~~
+ \frac{27}{64} \zeta_3 \alpha^4
+ \frac{243}{32} \zeta_4 \alpha^2
+ \frac{513}{256} \alpha^4
+ \frac{945}{64} \zeta_5 \alpha^3
+ \frac{2349}{128} \zeta_3 \alpha^2
+ \frac{3355}{2} \zeta_5 \Nf
\right. \nonumber \\
&& \left. ~~~~
+ \frac{8955}{16} \zeta_4 \Nf
+ \frac{9585}{256} \alpha^3
+ \frac{42255}{256} \zeta_5 \alpha^2
+ \frac{78075}{256} \alpha^2
+ \frac{247725}{128} \zeta_3 \alpha
- 33 \zeta_4 \Nf^2
\right. \nonumber \\
&& \left. ~~~~
+ \frac{1012023}{256} \zeta_3
+ \frac{2174765}{768} \alpha
- 156 \zeta_3 \Nf \alpha
+ 6 \zeta_3 \Nf^2 \alpha
\right] a^4 ~+~ O(a^5)
\nonumber \\
\left. \gamma_\alpha^{\MSbarss}(a,\alpha) \right|^{SU(3)} &=&
\left[
\frac{13}{2}
- \frac{3}{4} \alpha
- \frac{2}{3} \Nf
\right] a
+ \left[
\frac{531}{8}
- \frac{153}{16} \alpha
- \frac{61}{6} \Nf
- \frac{9}{16} \alpha^2
\right] a^2
\nonumber \\
&&
+ \left[
- \frac{12609}{128} \alpha
- \frac{8155}{36} \Nf
- \frac{837}{64} \alpha^2
- \frac{243}{16} \zeta_3
- \frac{81}{4} \zeta_3 \alpha
- \frac{81}{128} \alpha^3
+ \frac{153}{32} \Nf \alpha
\right. \nonumber \\
&& \left. ~~~~
+ \frac{215}{27} \Nf^2
+ \frac{29895}{32}
+ 33 \zeta_3 \Nf
\right] a^3
\nonumber \\
&&
+ \left[
\frac{10596127}{768}
- \frac{23350603}{5184} \Nf
- \frac{1012023}{256} \zeta_3
- \frac{204687}{128} \zeta_3 \alpha
- \frac{192377}{96} \alpha
\right. \nonumber \\
&& \left. ~~~~
- \frac{11853}{512} \alpha^3
- \frac{8955}{16} \zeta_4 \Nf
- \frac{6345}{32} \zeta_5 \alpha^2
- \frac{3355}{2} \zeta_5 \Nf
- \frac{3033}{16} \alpha^2
- \frac{513}{512} \alpha^4
\right. \nonumber \\
&& \left. ~~~~
- \frac{405}{128} \zeta_3 \alpha^3
- \frac{27}{128} \zeta_3 \alpha^4
- \frac{9}{8} \zeta_4 \Nf \alpha
- \frac{8}{3} \zeta_3 \Nf^3
+ \frac{237}{64} \Nf \alpha^2
+ \frac{727}{216} \Nf^2 \alpha
\right. \nonumber \\
&& \left. ~~~~
+ \frac{1161}{32} \zeta_3 \alpha^2
+ \frac{1833}{16} \zeta_3 \Nf \alpha
+ \frac{2017}{81} \zeta_3 \Nf^2
+ \frac{4427}{1458} \Nf^3
+ \frac{8019}{32} \zeta_4
+ \frac{33615}{128} \zeta_4 \alpha
\right. \nonumber \\
&& \left. ~~~~
+ \frac{40905}{4} \zeta_5
+ \frac{43033}{162} \Nf^2
+ \frac{119441}{576} \Nf \alpha
+ \frac{249345}{256} \zeta_5 \alpha
+ \frac{387649}{216} \zeta_3 \Nf
\right. \nonumber \\
&& \left. ~~~~
- 3 \zeta_3 \Nf^2 \alpha
+ 33 \zeta_4 \Nf^2
\right] a^4 ~+~ O(a^5) 
\nonumber \\
\left. \gamma_c^{\MSbarss}(a,\alpha) \right|^{SU(3)} &=&
\left[
\frac{3}{4} \alpha
- \frac{9}{4}
\right] a
+ \left[
\frac{5}{4} \Nf
- \frac{285}{16}
- \frac{9}{16} \alpha
+ \frac{9}{16} \alpha^2
\right] a^2
\nonumber \\
&&
+ \left[
\frac{637}{24} \Nf
- \frac{15817}{64}
- \frac{243}{32} \zeta_3
- \frac{81}{8} \zeta_3 \alpha
- \frac{63}{16} \Nf \alpha
+ \frac{33}{2} \zeta_3 \Nf
+ \frac{35}{36} \Nf^2
+ \frac{81}{128} \alpha^3
\right. \nonumber \\
&& \left. ~~~~
+ \frac{459}{32} \alpha
+ \frac{1485}{256} \alpha^2
\right] a^3
\nonumber \\
&&
+ \left[
\frac{1239661}{1152} \Nf
- \frac{2857419}{512}
- \frac{1924407}{512} \zeta_3
- \frac{76275}{512} \zeta_5 \alpha^2
- \frac{75573}{256} \zeta_3 \alpha
\right. \nonumber \\
&& \left. ~~~~
- \frac{12015}{256} \zeta_5 \alpha
- \frac{11107}{288} \Nf \alpha
- \frac{8955}{32} \zeta_4 \Nf
- \frac{3355}{4} \zeta_5 \Nf
- \frac{1215}{128} \zeta_3 \alpha^3
\right. \nonumber \\
&& \left. ~~~~
- \frac{779}{432} \Nf^2 \alpha
- \frac{586}{27} \Nf^2
- \frac{561}{64} \Nf \alpha^2
- \frac{153}{8} \zeta_4 \Nf \alpha
- \frac{135}{32} \zeta_3 \Nf \alpha^2
- \frac{55}{2} \zeta_3 \Nf^2
\right. \nonumber \\
&& \left. ~~~~
- \frac{39}{8} \zeta_3 \Nf \alpha
- \frac{4}{3} \zeta_3 \Nf^3
+ \frac{27}{128} \zeta_3 \alpha^4
+ \frac{33}{2} \zeta_4 \Nf^2
+ \frac{83}{108} \Nf^3
+ \frac{513}{512} \alpha^4
+ 3 \zeta_3 \Nf^2 \alpha
\right. \nonumber \\
&& \left. ~~~~
+ \frac{729}{64} \zeta_4 \alpha^2
+ \frac{2835}{128} \zeta_5 \alpha^3
+ \frac{5049}{512} \alpha^3
+ \frac{8019}{64} \zeta_4
+ \frac{17901}{128} \zeta_4 \alpha
+ \frac{25623}{256} \zeta_3 \alpha^2
\right. \nonumber \\
&& \left. ~~~~
+ \frac{40113}{512} \alpha^2
+ \frac{40905}{8} \zeta_5
+ \frac{48857}{48} \zeta_3 \Nf
+ \frac{368231}{1536} \alpha
\right] a^4 ~+~ O(a^5)
\nonumber \\
\left. \gamma_\psi^{\MSbarss}(a,\alpha) \right|^{SU(3)} &=&
\frac{4}{3} \alpha a
+ \left[
\frac{67}{3}
+ 8 \alpha
- \frac{4}{3} \Nf
\right] a^2
\nonumber \\
&&
+ \left[
\frac{20729}{36}
- \frac{550}{9} \Nf
- \frac{79}{2} \zeta_3
- \frac{17}{2} \Nf \alpha
+ \frac{9}{8} \alpha^3
+ \frac{20}{27} \Nf^2
+ \frac{45}{4} \alpha^2
+ \frac{789}{8} \alpha
\right. \nonumber \\
&& \left. ~~~~
+ 9 \zeta_3 \alpha
\right] a^3
\nonumber \\
&&
+ \left[
\frac{2109389}{162}
- \frac{761525}{1296} \zeta_5
- \frac{565939}{324} \zeta_3
- \frac{162103}{81} \Nf
- \frac{147541}{648} \Nf \alpha
\right. \nonumber \\
&& \left. ~~~~
- \frac{2291}{27} \zeta_3 \Nf
- \frac{1076}{243} \Nf^2 \alpha
- \frac{873}{8} \zeta_4 \alpha
- \frac{314}{3} \zeta_3 \Nf \alpha
- \frac{160}{3} \zeta_5 \Nf
- \frac{79}{2} \zeta_4 \Nf
\right. \nonumber \\
&& \left. ~~~~
- \frac{3}{2} \zeta_3 \alpha^3
- \frac{1}{8} \zeta_3 \alpha^4
+ \frac{3}{8} \alpha^4
+ \frac{16}{3} \zeta_3 \Nf^2 \alpha
+ \frac{81}{8} \zeta_4 \alpha^2
+ \frac{121}{2} \zeta_3 \alpha^2
+ \frac{291}{16} \alpha^3
\right. \nonumber \\
&& \left. ~~~~
+ \frac{140}{243} \Nf^3
+ \frac{160}{9} \zeta_3 \Nf^2
+ \frac{2607}{4} \zeta_4
+ \frac{3853}{81} \Nf^2
+ \frac{3933}{32} \alpha^2
+ \frac{8743}{8} \zeta_3 \alpha
+ \Nf \alpha^2
\right. \nonumber \\
&& \left. ~~~~
+ \frac{1640579}{864} \alpha
- 920 \zeta_5 \alpha
- 16 \zeta_4 \Nf \alpha
- 6 \zeta_3 \Nf \alpha^2
- 5 \zeta_5 \alpha^2
\right] a^4 ~+~ O(a^5) 
\end{eqnarray}
where $\zeta_n$ is the Riemann zeta function and when a quantity is labelled by
a scheme then the coupling constant and gauge parameter are in the same scheme.
For reasons of space we have provided the results for the $SU(3)$ group but the
expressions for an abritrary colour group are recorded in Appendix A. It is
evident there that the sum of $\gamma_A^{\MSbarss}(a,\alpha)$ and
$\gamma_\alpha^{\MSbarss}(a,\alpha)$ vanishes at four loops when 
$\alpha$~$=$~$0$ which acts as a minor check. The three loop parts of the 
anomalous dimensions are in complete agreement with earlier work \cite{31}. In 
addition we have determined the anomalous dimension of ${\cal O}$ directly by 
evaluating $G_{\cal O}^{ab}(p)$ to four loops and deducing the $\MSbar$ 
expression for $\gamma_{\cal O}(a,\alpha)$. We have checked that it satisfies 
the second Slavnov-Taylor identity of (\ref{gammastid}) explicitly to four 
loops. In determining $Z_{\cal O}$ from $G_{\cal O}^{ab}(p)$ we have also 
extracted $Z_g$ at four loops. While the four loop $\MSbar$ $\beta$-function 
was originally established in \cite{42} we have used the replacement 
(\ref{zozgreln}) to find $Z_g$ thereby reproducing the previously established
result. This was carried out for a {\em nonzero} $\alpha$. Although it may 
appear from (\ref{zozgreln}) that $Z_A$ is not required to find $Z_g$ by this 
method, and therefore contradict the first relation of (\ref{gammastid}), it is
in fact needed. This is because the tree term of $G_{\cal O}^{ab}(p)$ will 
involve $\alpha$ from the ghost term in the definition of ${\cal O}$ when the 
operator is inserted in a ghost $2$-point function. Then $Z_A$ arises from the 
gauge parameter mapping of (\ref{zdef}). The other motivation for extracting 
$Z_g$ in this manner is to test its viability as an alternative way of finding 
the $\beta$-function without the need to evaluate a vertex function.

\sect{$\mMOM$ scheme results.}

Having established our four loop setup for (\ref{lagcf}) in the $\MSbar$ scheme
we can now carry out the renormalization in another scheme which is the 
minimal MOM scheme. It was introduced in \cite{48} for the linear covariant
gauge and is constructed from the Taylor theorem property that the ghost-gluon
vertex is finite in the Landau gauge. To assist with streamlining lattice
field theory computations the coupling renormalization constant in the
$\mMOM$ scheme is defined by a condition that seeks to preserve properties of
the Taylor nonrenormalization theorem for nonzero gauge parameters. In other
words the starting point used in \cite{48} is (\ref{zgccg}) but without 
$Z_\alpha$ which would be the Taylor identity in the Landau gauge. It led to 
the $\mMOM$ scheme definition given in \cite{48} which is
\begin{equation}
\left. Z_g^{\MSbarss} \sqrt{Z_A^{\MSbarss}} Z_c^{\MSbarss} \right|^{\linss}~=~ 
\left. Z_g^{\mMOMss} \sqrt{Z_A^{\mMOMss}} Z_c^{\mMOMss} \right|^{\linss} 
\label{linmmomdef}
\end{equation}
for all $\alpha$. In this relation $Z_g^{\mMOMss}$ is deduced from the known 
$\MSbar$ expressions for $Z_g$, $Z_A$ and $Z_c$ while the $\mMOM$ counterparts 
of the ghost and gluon are determined by ensuring that after renormalization 
the respective $2$-point functions have no $O(a)$ terms with the quark field
being renormalized in the same way too. This leaves $Z_g^{\mMOMss}$ as the only 
unknown. However in extracting the expression for it iteratively the $\mMOM$ 
coupling constant, $a_{\mMOMss}$, has to be expressed as a perturbative 
expansion in $a_{\MSbarss}$. Otherwise the condition would have poles in 
$\epsilon$. Using (\ref{linmmomdef}) the four loop $\mMOM$ $\beta$-function was
deduced in \cite{48}. This was extended to five loops in \cite{58} for 
$\alpha$~$\neq$~$0$.

Given this overview of the $\mMOM$ scheme for linear covariant gauges there is 
a natural extension of this scheme for the Curci-Ferrari gauge based on the
parallel Slavnov-Taylor identity. In other words the scheme is defined by the
condition
\begin{equation}
Z_g^{\MSbarss} \sqrt{Z_A^{\MSbarss}} Z_c^{\MSbarss} 
\left( Z_\alpha^{\MSbarss} \right)^{-2} ~=~ 
Z_g^{\mMOMss} \sqrt{Z_A^{\mMOMss}} Z_c^{\mMOMss} 
\left( Z_\alpha^{\mMOMss} \right)^{-2}
\label{cfmmomdef}
\end{equation}
for all $\alpha$. The validity for all $\alpha$ indicates a subtle difference
between the definition of the scheme in different gauges. In the linear 
covariant gauge case the motivation was to preserve a property of one 
particular gauge parameter choice for all values. The nature of (\ref{zgccg}) 
is such that (\ref{cfmmomdef}) is already valid for all $\alpha$. Moreover the 
combination of renormalization constants in both schemes is unity by 
construction. So no coupling constant mapping is actually required to deduce 
$Z_g^{\mMOMss}$. All that is needed is the extraction of the ghost, gluon and 
gauge parameter renormalization constants from the $2$-point functions under 
the renormalization condition that there are no $O(a)$ corrections. Then the
coupling renormalization constant is determined from setting the right hand 
side of (\ref{cfmmomdef}) to unity.

From the $2$-point functions that were evaluated to determine the $\MSbar$
results of the previous section it is straightforward to extract the $\mMOM$
renormalization group functions for the field anomalous dimensions to four
loops. For $SU(3)$ we have
\begin{eqnarray}
\left. \gamma_A^{\mMOMss}(a,\alpha) \right|^{SU(3)} &=&
\left[
\frac{3}{2} \alpha
+ \frac{2}{3} \Nf
- \frac{13}{2}
\right] a
\nonumber \\
&&
+ \left[
\frac{67}{6} \Nf
- \frac{255}{4}
- \frac{9}{8} \alpha^3
- \frac{3}{8} \alpha^2
+ \frac{363}{8} \alpha
- 5 \Nf \alpha
- \Nf \alpha^2
\right] a^2
\nonumber \\
&&
+ \left[
324 \zeta_3
- \frac{8637}{4}
- \frac{7089}{32} \alpha^2
- \frac{2565}{16} \zeta_3 \alpha
- \frac{1397}{8} \Nf \alpha
- \frac{945}{32} \alpha^3
- \frac{719}{54} \Nf^2
\right. \nonumber \\
&& \left. ~~~~
- \frac{229}{12} \zeta_3 \Nf
- \frac{45}{4} \zeta_3 \alpha^2
- \frac{27}{16} \zeta_3 \alpha^3
- \frac{8}{9} \zeta_3 \Nf^2
- \frac{3}{4} \zeta_3 \Nf \alpha^2
+ \frac{3}{8} \Nf \alpha^4
\right. \nonumber \\
&& \left. ~~~~
+ \frac{9}{2} \zeta_3 \Nf \alpha
+ \frac{39}{4} \Nf \alpha^3
+ \frac{89}{8} \Nf \alpha^2
+ \frac{387}{64} \alpha^4
+ \frac{4005}{4} \alpha
+ \frac{11227}{24} \Nf
\right] a^3
\nonumber \\
&&
+ \left[
3312 \zeta_3 \Nf \alpha
- \frac{27189875}{256}
- \frac{3070587}{128} \zeta_3 \alpha
- \frac{1902141}{256} \alpha^2
- \frac{1118977}{648} \Nf^2
\right. \nonumber \\
&& \left. ~~~~
- \frac{921265}{144} \zeta_5 \Nf
- \frac{889231}{144} \zeta_3 \Nf
- \frac{364619}{32} \Nf \alpha
- \frac{7263}{256} \alpha^5
- \frac{2295}{8} \zeta_5 \Nf \alpha
\right. \nonumber \\
&& \left. ~~~~
- \frac{1983}{32} \Nf \alpha^4
- \frac{405}{16} \zeta_5 \alpha^4
- \frac{45}{4} \zeta_5 \Nf \alpha^2
- \frac{16}{9} \zeta_3 \Nf^3
- \frac{15}{2} \zeta_5 \Nf \alpha^3
\right. \nonumber \\
&& \left. ~~~~
- \frac{9}{2} \Nf \alpha^5
- \frac{8}{9} \Nf^3 \alpha
- \frac{4}{3} \zeta_3 \Nf^2 \alpha^2
+ \frac{9}{8} \zeta_3 \Nf \alpha^4
+ \frac{15}{8} \zeta_3 \Nf \alpha^3
+ \frac{32}{9} \zeta_3 \Nf^3 \alpha
\right. \nonumber \\
&& \left. ~~~~
+ \frac{61}{16} \zeta_3 \Nf \alpha^2
+ \frac{81}{64} \zeta_3 \alpha^5
+ \frac{665}{27} \Nf^3
+ \frac{765}{16} \zeta_5 \alpha^3
+ \frac{1028}{3} \Nf^2 \alpha
+ \frac{1339}{36} \Nf^2 \alpha^2
\right. \nonumber \\
&& \left. ~~~~
+ \frac{1971}{32} \zeta_3 \alpha^4
+ \frac{5143}{27} \zeta_3 \Nf^2
+ \frac{5517}{16} \alpha^3
+ \frac{6195}{32} \Nf \alpha^3
+ \frac{9280}{27} \zeta_5 \Nf^2
\right. \nonumber \\
&& \left. ~~~~
+ \frac{11403}{8} \zeta_3 \alpha^2
+ \frac{46335}{64} \Nf \alpha^2
+ \frac{79965}{256} \alpha^4
+ \frac{81657}{128} \zeta_3 \alpha^3
+ \frac{459135}{256} \zeta_5 \alpha^2
\right. \nonumber \\
&& \left. ~~~~
+ \frac{879525}{128} \zeta_5 \alpha
+ \frac{1950705}{128} \zeta_5
+ \frac{5549393}{192} \Nf
+ \frac{7740879}{256} \zeta_3
+ \frac{13104261}{256} \alpha
\right. \nonumber \\
&& \left. ~~~~
- 161 \zeta_3 \Nf^2 \alpha
\right] a^4 ~+~ O(a^5)
\nonumber \\
\left. \gamma_\alpha^{\mMOMss}(a,\alpha) \right|^{SU(3)} &=&
\left[
\frac{13}{2}
- \frac{3}{4} \alpha
- \frac{2}{3} \Nf
\right] a
\nonumber \\
&&
+ \left[
5 \Nf \alpha
- \frac{645}{16} \alpha
- \frac{67}{6} \Nf
- \frac{21}{16} \alpha^2
+ \frac{9}{8} \alpha^3
+ \frac{255}{4}
+ \Nf \alpha^2
\right] a^2
\nonumber \\
&&
+ \left[
\frac{8637}{4}
- \frac{14013}{16} \alpha
- \frac{11227}{24} \Nf
- \frac{387}{64} \alpha^4
- \frac{89}{8} \Nf \alpha^2
- \frac{39}{4} \Nf \alpha^3
\right. \nonumber \\
&& \left. ~~~~
- \frac{9}{4} \zeta_3 \Nf \alpha
- \frac{3}{8} \Nf \alpha^4
+ \frac{3}{4} \zeta_3 \Nf \alpha^2
+ \frac{8}{9} \zeta_3 \Nf^2
+ \frac{27}{16} \zeta_3 \alpha^3
+ \frac{45}{4} \zeta_3 \alpha^2
\right. \nonumber \\
&& \left. ~~~~
+ \frac{229}{12} \zeta_3 \Nf
+ \frac{719}{54} \Nf^2
+ \frac{1053}{32} \alpha^3
+ \frac{1355}{8} \Nf \alpha
+ \frac{3861}{32} \zeta_3 \alpha
+ \frac{5631}{32} \alpha^2
\right. \nonumber \\
&& \left. ~~~~
- 324 \zeta_3
\right] a^3
\nonumber \\
&&
+ \left[
\frac{27189875}{256}
- \frac{11625507}{256} \alpha
- \frac{7740879}{256} \zeta_3
- \frac{5549393}{192} \Nf
- \frac{1950705}{128} \zeta_5
\right. \nonumber \\
&& \left. ~~~~
- \frac{1571535}{256} \zeta_5 \alpha
- \frac{156933}{512} \alpha^4
- \frac{109629}{128} \zeta_3 \alpha^2
- \frac{82125}{512} \alpha^3
- \frac{73665}{128} \zeta_3 \alpha^3
\right. \nonumber \\
&& \left. ~~~~
- \frac{57645}{32} \zeta_5 \alpha^2
- \frac{40359}{64} \Nf \alpha^2
- \frac{9280}{27} \zeta_5 \Nf^2
- \frac{6291}{2} \zeta_3 \Nf \alpha
- \frac{5943}{32} \Nf \alpha^3
\right. \nonumber \\
&& \left. ~~~~
- \frac{5143}{27} \zeta_3 \Nf^2
- \frac{3953}{12} \Nf^2 \alpha
- \frac{2925}{64} \zeta_5 \alpha^3
- \frac{1339}{36} \Nf^2 \alpha^2
- \frac{665}{27} \Nf^3
\right. \nonumber \\
&& \left. ~~~~
- \frac{243}{4} \zeta_3 \alpha^4
- \frac{157}{4} \zeta_3 \Nf \alpha^2
- \frac{81}{64} \zeta_3 \alpha^5
- \frac{32}{9} \zeta_3 \Nf^3 \alpha
- \frac{21}{4} \zeta_3 \Nf \alpha^3
\right. \nonumber \\
&& \left. ~~~~
- \frac{9}{8} \zeta_3 \Nf \alpha^4
+ \frac{4}{3} \zeta_3 \Nf^2 \alpha^2
+ \frac{8}{9} \Nf^3 \alpha
+ \frac{9}{2} \Nf \alpha^5
+ \frac{15}{2} \zeta_5 \Nf \alpha^3
+ \frac{16}{9} \zeta_3 \Nf^3
\right. \nonumber \\
&& \left. ~~~~
+ \frac{45}{4} \zeta_5 \Nf \alpha^2
+ \frac{319}{2} \zeta_3 \Nf^2 \alpha
+ \frac{405}{16} \zeta_5 \alpha^4
+ \frac{945}{4} \zeta_5 \Nf \alpha
+ \frac{1983}{32} \Nf \alpha^4
\right. \nonumber \\
&& \left. ~~~~
+ \frac{7263}{256} \alpha^5
+ \frac{86411}{8} \Nf \alpha
+ \frac{168003}{32} \alpha^2
+ \frac{889231}{144} \zeta_3 \Nf
+ \frac{921265}{144} \zeta_5 \Nf
\right. \nonumber \\
&& \left. ~~~~
+ \frac{1118977}{648} \Nf^2
+ \frac{2786007}{128} \zeta_3 \alpha
\right] a^4 ~+~ O(a^5)
\nonumber \\
\left. \gamma_c^{\mMOMss}(a,\alpha) \right|^{SU(3)} &=&
\left[
\frac{3}{4} \alpha
- \frac{9}{4}
\right] a
+ \left[
\frac{3}{4} \Nf
- \frac{153}{8}
- \frac{9}{8} \alpha^2
+ \frac{189}{16} \alpha
\right] a^2
\nonumber \\
&&
+ \left[
\frac{1401}{16} \Nf
- \frac{11691}{16}
- \frac{6399}{128} \alpha^2
- \frac{783}{32} \zeta_3 \alpha
- \frac{201}{16} \Nf \alpha
- \frac{9}{8} \Nf \alpha^2
- \frac{5}{2} \Nf^2
\right. \nonumber \\
&& \left. ~~~~
+ \frac{9}{4} \zeta_3 \Nf
+ \frac{9}{4} \zeta_3 \Nf \alpha
+ \frac{27}{64} \alpha^3
+ \frac{81}{32} \zeta_3 \alpha^2
+ \frac{1269}{16} \zeta_3
+ \frac{20691}{64} \alpha
\right] a^3
\nonumber \\
&&
+ \left[
26 \zeta_3 \Nf^2
- \frac{16985133}{512}
- \frac{880173}{256} \zeta_3 \alpha
- \frac{568647}{256} \alpha^2
- \frac{229095}{512} \zeta_5 \alpha^2
\right. \nonumber \\
&& \left. ~~~~
- \frac{62481}{32} \Nf \alpha
- \frac{16059}{32} \zeta_3 \Nf
- \frac{13115}{48} \Nf^2
- \frac{5895}{16} \zeta_5 \Nf
- \frac{5787}{128} \Nf \alpha^2
\right. \nonumber \\
&& \left. ~~~~
- \frac{1287}{32} \zeta_3 \Nf \alpha^2
- \frac{585}{8} \zeta_5 \Nf \alpha
- \frac{405}{128} \zeta_3 \alpha^3
- \frac{135}{128} \zeta_3 \alpha^4
- \frac{27}{8} \zeta_3 \Nf \alpha^3
\right. \nonumber \\
&& \left. ~~~~
+ \frac{5}{2} \Nf^3
+ \frac{27}{64} \Nf \alpha^4
+ \frac{45}{8} \Nf^2 \alpha^2
+ \frac{45}{8} \zeta_5 \Nf \alpha^2
+ \frac{171}{8} \Nf \alpha^3
+ \frac{409}{8} \Nf^2 \alpha
\right. \nonumber \\
&& \left. ~~~~
+ \frac{1701}{128} \alpha^4
+ \frac{2241}{512} \alpha^3
+ \frac{2835}{128} \zeta_5 \alpha^3
+ \frac{47007}{128} \zeta_3 \alpha^2
+ \frac{181845}{256} \zeta_5 \alpha
\right. \nonumber \\
&& \left. ~~~~
+ \frac{739053}{128} \Nf
+ \frac{1140075}{256} \zeta_5
+ \frac{2264193}{512} \zeta_3
+ \frac{8902611}{512} \alpha
+ 6 \zeta_3 \Nf^2 \alpha
\right. \nonumber \\
&& \left. ~~~~
+ 9 \zeta_3 \Nf \alpha
\right] a^4 ~+~ O(a^5) 
\nonumber \\
\left. \gamma_\psi^{\mMOMss}(a,\alpha) \right|^{SU(3)} &=&
\frac{4}{3} \alpha a
+ \left[
\frac{67}{3}
- \frac{4}{3} \Nf
- 5 \alpha^2
+ 6 \alpha
\right] a^2
\nonumber \\
&&
+ \left[
16 \zeta_3 \Nf
- \frac{706}{9} \Nf
- \frac{645}{4} \alpha
- \frac{607}{2} \zeta_3
- \frac{421}{4} \alpha^2
- \frac{309}{2} \zeta_3 \alpha
- \frac{9}{2} \zeta_3 \alpha^2
\right. \nonumber \\
&& \left. ~~~~
+ \frac{8}{9} \Nf^2
+ \frac{40}{3} \Nf \alpha
+ \frac{245}{12} \alpha^3
+ \frac{29675}{36}
+ 2 \Nf \alpha^2
+ 8 \zeta_3 \Nf \alpha
\right] a^3
\nonumber \\
&&
+ \left[
\frac{31003343}{648}
- \frac{21683117}{648} \zeta_3
- \frac{2393555}{324} \Nf
- \frac{986453}{72} \alpha
- \frac{88873}{48} \alpha^2
\right. \nonumber \\
&& \left. ~~~~
- \frac{9095}{144} \alpha^4
- \frac{1640}{9} \zeta_5 \Nf \alpha
- \frac{536}{3} \zeta_3 \Nf \alpha^2
- \frac{272}{9} \zeta_3 \Nf^2
- \frac{244}{27} \Nf^2 \alpha
\right. \nonumber \\
&& \left. ~~~~
- \frac{49}{2} \Nf \alpha^3
- \frac{45}{2} \zeta_5 \alpha^3
- \frac{40}{9} \Nf^3
- \frac{3}{4} \Nf \alpha^4
+ \frac{43}{36} \zeta_3 \alpha^4
+ \frac{226}{3} \zeta_3 \alpha
\right. \nonumber \\
&& \left. ~~~~
+ \frac{619}{6} \Nf \alpha^2
+ \frac{735}{8} \zeta_5 \alpha^2
+ \frac{896}{9} \zeta_3 \Nf \alpha
+ \frac{2861}{9} \Nf^2
+ \frac{3497}{12} \zeta_3 \alpha^3
\right. \nonumber \\
&& \left. ~~~~
+ \frac{12143}{12} \alpha^3
+ \frac{34175}{12} \zeta_5 \alpha
+ \frac{50693}{16} \zeta_3 \alpha^2
+ \frac{68183}{54} \Nf \alpha
+ \frac{74440}{27} \zeta_3 \Nf
\right. \nonumber \\
&& \left. ~~~~
+ \frac{15846715}{1296} \zeta_5
- 830 \zeta_5 \Nf
- 12 \zeta_3 \Nf \alpha^3
- 2 \Nf^2 \alpha^2
\right] a^4 ~+~ O(a^5) ~.
\end{eqnarray}
We have provided the expressions for $\alpha$~$\neq$~$0$ as those for 
$\alpha$~$=$~$0$ are in full agreement with the field anomalous dimensions of
\cite{58,59,60}. This acts as a partial check on our determination.

Although the $\alpha$~$\neq$~$0$ four loop $\mMOM$ $\beta$-function will not
agree with the corresponding linear covariant gauge result of \cite{48,58},
rather than recording it at this juncture we can in fact determine it at 
{\em five} loops. This is made possible through the renormalization group. When
a renormalization group function for one quantity is known at $L$ loops in one 
scheme and $(L+1)$ loops in another then a conversion function can be 
constructed based on the $L$ loop renormalization constants of the underlying 
quantities in the separate schemes. The $L$ loop conversion function is then 
used in a renormalization group context to access the unknown $(L+1)$ loop
renormalization group function. More specifically for an element of the set
$\phi$ defined earlier the conversion function $C_\phi(a,\alpha)$ is given by
\begin{equation}
C_\phi(a_{\MSbarss},\alpha_{\MSbarss}) ~=~ 
\frac{Z^{\mMOMss}_\phi}{Z^{\MSbarss}_\phi}
\end{equation}
where the variables of $Z^{\mMOMss}_\phi$ have to be iteratively mapped to 
those of the $\MSbar$ scheme via
\begin{equation}
g_{\mMOMss} ~=~ \frac{Z_g^{\MSbarss}}{Z_g^{\mMOMss}} \, g_{\MSbarss} ~~~,~~~
\alpha_{\mMOMss} ~=~ \frac{Z_A^{\mMOMss}}{Z_A^{\MSbarss}} \alpha_{\MSbarss} ~.
\end{equation}
When these mappings have been computed to the loop order that the 
renormalization constants are known to then the respective renormalization
group functions of the scheme not available at $(L+1)$ loops, which in our case
is the $\mMOM$ scheme, are deduced from 
\begin{eqnarray}
\beta^{\mbox{$\mMOMss$}} ( a_{\mbox{$\mMOMss$}}, \alpha_{\mbox{$\mMOMss$}} ) &=&
\left[ \beta^{\mbox{$\MSbarss$}}( a_{\mbox{$\MSbarss$}} )
\frac{\partial a_{\mbox{$\mMOMss$}}}{\partial a_{\mbox{$\MSbarss$}}} 
~+~ \alpha_{\mbox{$\MSbarss$}} \gamma^{\mbox{$\MSbarss$}}_\alpha
( a_{\mbox{$\MSbarss$}}, \alpha_{\mbox{\footnotesize{$\MSbarss$}}} )
\frac{\partial a_{\mbox{$\mMOMss$}}}{\partial \alpha_{\mbox{$\MSbarss$}}}
\right]_{\stackrel{\MSbars \to}{\mMOMss}} ~~~~~~
\end{eqnarray}
and
\begin{eqnarray}
\gamma^{\mMOMss}_\phi \left( a_{\mMOMss}, \alpha_{\mMOMss} \right) &=& \left[
\gamma^{\MSbarss}_\phi \left( a_{\MSbarss}, \alpha_{\MSbarss} \right) +
\beta^{\MSbarss} \left( a_{\MSbarss} \right)
\frac{\partial ~}{\partial a_{\MSbarss}}
\ln C_\phi \left( a_{\MSbarss}, \alpha_{\MSbarss} \right) \right. \nonumber \\
&& \left. \,+\, \alpha_{\MSbarss} \,
\gamma^{\MSbarss}_\alpha \left( a_{\MSbarss}, \alpha_{\MSbarss} \right)
\frac{\partial ~}{\partial \alpha_{\MSbarss}}
\ln C_\phi \left( a_{\MSbarss}, \alpha_{\MSbarss} \right)
\right]_{\stackrel{\MSbars \to}{\mMOMss}} ~.~~~~~ 
\end{eqnarray}
The scheme map on the right side indicates that the $\MSbar$ variables have to 
mapped to the $\mMOM$ scheme to be consistent with the arguments of the
functions on the left side. 

Following this prescription the $SU(3)$ five loop $\mMOM$ $\beta$-function for 
$\alpha$~$\neq$~$0$ is 
\begin{eqnarray}
\left. \beta^{\mMOMss}(a,\alpha) \right|^{SU(3)} &=&
 \left[
\frac{2}{3} \Nf
- 11
\right] a^2
\nonumber \\
&&
+ \left[
\frac{38}{3} \Nf
- \frac{9}{8} \alpha^3
+ \frac{33}{8} \alpha^2
+ \frac{195}{4} \alpha
- 102
- 5 \Nf \alpha
- \Nf \alpha^2
\right] a^3
\nonumber \\
&&
+ \left[
\frac{7715}{12} \Nf
- \frac{28965}{8}
- \frac{8913}{64} \alpha^2
- \frac{989}{54} \Nf^2
- \frac{715}{4} \Nf \alpha
- \frac{675}{16} \alpha^3
\right. \nonumber \\
&& \left. ~~~~
- \frac{405}{8} \zeta_3 \alpha
- \frac{175}{12} \zeta_3 \Nf
- \frac{99}{16} \zeta_3 \alpha^2
- \frac{27}{16} \zeta_3 \alpha^3
- \frac{8}{9} \zeta_3 \Nf^2
- \frac{3}{4} \zeta_3 \Nf \alpha^2
\right. \nonumber \\
&& \left. ~~~~
+ \frac{3}{8} \Nf \alpha^4
+ \frac{39}{4} \Nf \alpha^3
+ \frac{71}{8} \Nf \alpha^2
+ \frac{387}{64} \alpha^4
+ \frac{3861}{8} \zeta_3
+ \frac{36675}{32} \alpha
\right] a^4
\nonumber \\
&&
+ \left[
\frac{970819}{24} \Nf
- \frac{1380469}{8}
- \frac{1027375}{144} \zeta_5 \Nf
- \frac{806967}{256} \alpha^2
- \frac{736541}{324} \Nf^2
\right. \nonumber \\
&& \left. ~~~~
- \frac{516881}{72} \zeta_3 \Nf
- \frac{413681}{32} \Nf \alpha
- \frac{351555}{16} \zeta_3 \alpha
- \frac{98325}{256} \alpha^3
\right. \nonumber \\
&& \left. ~~~~
- \frac{7407}{64} \zeta_3 \alpha^2
- \frac{7263}{256} \alpha^5
- \frac{1845}{8} \zeta_5 \Nf \alpha
- \frac{489}{8} \Nf \alpha^4
- \frac{405}{16} \zeta_5 \alpha^4
\right. \nonumber \\
&& \left. ~~~~
- \frac{16}{9} \zeta_3 \Nf^3
- \frac{15}{2} \zeta_5 \Nf \alpha^3
- \frac{9}{2} \Nf \alpha^5
- \frac{8}{9} \Nf^3 \alpha
- \frac{4}{3} \zeta_3 \Nf^2 \alpha^2
+ \frac{9}{8} \zeta_3 \Nf \alpha^4
\right. \nonumber \\
&& \left. ~~~~
+ \frac{32}{9} \zeta_3 \Nf^3 \alpha
+ \frac{69}{8} \zeta_3 \Nf \alpha^3
+ \frac{81}{64} \zeta_3 \alpha^5
+ \frac{436}{9} \Nf^2 \alpha^2
+ \frac{521}{8} \zeta_3 \Nf \alpha^2
\right. \nonumber \\
&& \left. ~~~~
+ \frac{800}{27} \Nf^3
+ \frac{3591}{64} \zeta_3 \alpha^4
+ \frac{4161}{16} \Nf \alpha^2
+ \frac{4703}{12} \Nf^2 \alpha
+ \frac{5355}{64} \zeta_5 \alpha^3
\right. \nonumber \\
&& \left. ~~~~
+ \frac{6547}{27} \zeta_3 \Nf^2
+ \frac{6555}{32} \Nf \alpha^3
+ \frac{9280}{27} \zeta_5 \Nf^2
+ \frac{48879}{128} \zeta_3 \alpha^3
+ \frac{59535}{64} \zeta_5 \alpha^2
\right. \nonumber \\
&& \left. ~~~~
+ \frac{80775}{256} \alpha^4
+ \frac{171585}{32} \zeta_5 \alpha
+ \frac{625317}{16} \zeta_3
+ \frac{772695}{32} \zeta_5
+ \frac{1005741}{16} \alpha
\right. \nonumber \\
&& \left. ~~~~
- 143 \zeta_3 \Nf^2 \alpha
+ 2664 \zeta_3 \Nf \alpha
\right] a^5
\nonumber \\
&&
+ \left[
\frac{65264845}{324} \zeta_5 \Nf^2
- \frac{21619456551}{4096} \zeta_7
- \frac{18219328375}{6912} \zeta_5 \Nf
\right. \nonumber \\
&& \left. ~~~~
- \frac{10327103555}{20736} \zeta_3 \Nf
- \frac{3248220045}{256}
- \frac{2997550521}{8192} \zeta_7 \alpha
\right. \nonumber \\
&& \left. ~~~~
+ \frac{4922799165}{512} \zeta_5
+ \frac{24870449471}{18432} \zeta_7 \Nf
+ \frac{115659378547}{31104} \Nf
\right. \nonumber \\
&& \left. ~~~~
- \frac{1703225781}{1024} \zeta_3 \alpha
- \frac{833934985}{2592} \Nf^2
- \frac{701614431}{16384} \zeta_7 \alpha^2
\right. \nonumber \\
&& \left. ~~~~
- \frac{334971945}{2048} \zeta_5 \alpha^2
- \frac{287091579}{1024} \alpha^2
- \frac{67011685}{64} \Nf \alpha
\right. \nonumber \\
&& \left. ~~~~
- \frac{61995825}{1024} \alpha^3
- \frac{60018273}{8192} \zeta_7 \alpha^3
- \frac{33569371}{768} \zeta_3 \Nf \alpha^2
\right. \nonumber \\
&& \left. ~~~~
- \frac{26952037}{432} \zeta_7 \Nf^2
- \frac{26347275}{512} \zeta_3^2 \alpha
- \frac{21835305}{512} \zeta_5 \alpha^3
- \frac{11222217}{2048} \zeta_3 \alpha^4
\right. \nonumber \\
&& \left. ~~~~
- \frac{10853415}{2048} \zeta_5 \alpha^4
- \frac{9356067}{16384} \zeta_7 \alpha^4
- \frac{2277045}{1024} \zeta_3^2 \alpha^2
- \frac{1503875}{72} \zeta_3 \Nf^2 \alpha
\right. \nonumber \\
&& \left. ~~~~
- \frac{570465}{256} \Nf \alpha^4
- \frac{299875}{54} \zeta_5 \Nf^3
- \frac{294171}{32} \zeta_3 \Nf \alpha^3
- \frac{191585}{144} \zeta_5 \Nf^2 \alpha^2
\right. \nonumber \\
&& \left. ~~~~
- \frac{124443}{64} \alpha^5
- \frac{82869}{32} \zeta_3^2 \Nf
- \frac{59531}{36} \zeta_3^2 \Nf^2
- \frac{44290}{9} \zeta_5 \Nf^2 \alpha
\right. \nonumber \\
&& \left. ~~~~
- \frac{29883}{128} \zeta_3 \Nf \alpha^4
- \frac{19277}{16} \Nf^2 \alpha^3
- \frac{11745}{256} \zeta_3 \alpha^6
- \frac{8967}{8} \zeta_7 \Nf^2 \alpha
\right. \nonumber \\
&& \left. ~~~~
- \frac{5589}{4} \zeta_3 \alpha^5
- \frac{2617}{27} \Nf^4
- \frac{1663}{2} \Nf^3 \alpha
- \frac{1423}{18} \Nf^3 \alpha^2
- \frac{755}{16} \Nf^2 \alpha^4
\right. \nonumber \\
&& \left. ~~~~
- \frac{531}{32} \zeta_3 \Nf \alpha^5
- \frac{304}{27} \zeta_3 \Nf^4
- \frac{290}{3} \zeta_5 \Nf^3 \alpha
- \frac{32}{3} \zeta_3 \Nf^3 \alpha^3
- \frac{27}{8} \zeta_3^2 \Nf \alpha^3
\right. \nonumber \\
&& \left. ~~~~
- \frac{27}{64} \zeta_3 \Nf \alpha^6
- \frac{15}{4} \zeta_5 \Nf^2 \alpha^3
+ \frac{5}{3} \zeta_5 \Nf^3 \alpha^2
+ \frac{8}{3} \Nf^3 \alpha^3
+ \frac{9}{4} \zeta_3^2 \Nf^2 \alpha^2
\right. \nonumber \\
&& \left. ~~~~
+ \frac{45}{4} \zeta_5 \Nf \alpha^5
+ \frac{729}{128} \zeta_3^2 \Nf \alpha^4
+ \frac{1197}{32} \Nf \alpha^6
+ \frac{1215}{64} \zeta_5 \alpha^6
+ \frac{1760}{27} \zeta_5 \Nf^4
\right. \nonumber \\
&& \left. ~~~~
+ \frac{2240}{27} \zeta_3^2 \Nf^3
+ \frac{3865}{9} \zeta_3 \Nf^3 \alpha
+ \frac{4149}{32} \zeta_3^2 \Nf \alpha^2
+ \frac{6075}{256} \zeta_3^2 \alpha^5
\right. \nonumber \\
&& \left. ~~~~
+ \frac{6555}{32} \zeta_5 \Nf \alpha^4
+ \frac{11907}{512} \zeta_7 \Nf \alpha^4
+ \frac{30897}{128} \Nf \alpha^5
+ \frac{45225}{128} \zeta_5 \alpha^5
\right. \nonumber \\
&& \left. ~~~~
+ \frac{54795}{64} \zeta_5 \Nf \alpha^3
+ \frac{57869}{18} \zeta_3 \Nf^2 \alpha^2
+ \frac{107163}{1024} \zeta_7 \alpha^5
+ \frac{126117}{1024} \zeta_3^2 \alpha^4
\right. \nonumber \\
&& \left. ~~~~
+ \frac{129869}{162} \zeta_3 \Nf^3
+ \frac{221967}{2048} \alpha^6
+ \frac{283743}{512} \zeta_3^2 \alpha^3
+ \frac{285381}{64} \zeta_3^2 \Nf \alpha
\right. \nonumber \\
&& \left. ~~~~
+ \frac{438459}{256} \zeta_7 \Nf \alpha^2
+ \frac{1657439}{864} \Nf^2 \alpha^2
+ \frac{3249767}{324} \Nf^3
+ \frac{7436727}{128} \zeta_3 \alpha^3
\right. \nonumber \\
&& \left. ~~~~
+ \frac{7696161}{64} \zeta_3^2
+ \frac{12498839}{256} \Nf \alpha^2
+ \frac{13019053}{1296} \zeta_3 \Nf^2
+ \frac{16978059}{2048} \alpha^4
\right. \nonumber \\
&& \left. ~~~~
+ \frac{23484425}{768} \zeta_5 \Nf \alpha^2
+ \frac{26378669}{432} \Nf^2 \alpha
+ \frac{39900215}{384} \zeta_5 \Nf \alpha
\right. \nonumber \\
&& \left. ~~~~
+ \frac{41664567}{1024} \zeta_7 \Nf \alpha
+ \frac{55392705}{1024} \zeta_5 \alpha
+ \frac{129737865}{32} \alpha
- 78 \zeta_3^2 \Nf^2 \alpha
\right. \nonumber \\
&& \left. ~~~~
+ \frac{139712267}{384} \zeta_3 \Nf \alpha
+ \frac{685178523}{2048} \zeta_3 \alpha^2
+ \frac{1064190195}{512} \zeta_3
- 89 \zeta_3 \Nf^3 \alpha^2
\right. \nonumber \\
&& \left. ~~~~
+ 6 \zeta_3 \Nf^2 \alpha^4
+ 540 \zeta_3 \Nf^2 \alpha^3
+ 20364 \Nf \alpha^3
\right] a^6 ~+~ O(a^7) ~.
\label{betammom5}
\end{eqnarray}
As the anomalous dimension of the quark mass operator, $\bar{\psi}\psi$, is 
known to five loops in the $\MSbar$ scheme \cite{61,62,63,64,65,66,67} we can
also apply the same formalism and deduce the five loop $\mMOM$ anomalous 
dimension. For the same group we have
\begin{eqnarray}
\left. \gamma_{\bar{\psi}\psi}^{\mMOMss}(a,\alpha) \right|^{SU(3)} &=&
-~ 4 a
+ \left[
\frac{4}{3} \Nf
- \frac{209}{3}
+ 2 \alpha^2
+ 24 \alpha
\right] a^2
\nonumber \\
&&
+ \left[
\frac{4742}{27} \Nf
- \frac{95383}{36}
- \frac{176}{9} \zeta_3 \Nf
- \frac{93}{2} \alpha^2
- \frac{50}{3} \alpha^3
- \frac{8}{3} \Nf^2
+ \frac{9}{2} \zeta_3 \alpha^2
\right. \nonumber \\
&& \left. ~~~~
+ \frac{3949}{4} \alpha
+ \frac{5635}{6} \zeta_3
- 14 \Nf \alpha
- 2 \Nf \alpha^2
+ 45 \zeta_3 \alpha
\right] a^3
\nonumber \\
&&
+ \left[
\frac{5246557}{324} \Nf
- \frac{182707879}{1296}
- \frac{376037}{144} \zeta_3 \alpha^2
- \frac{309295}{48} \zeta_5
- \frac{159817}{27} \zeta_3 \Nf
\right. \nonumber \\
&& \left. ~~~~
- \frac{97915}{24} \alpha^2
- \frac{91399}{27} \Nf \alpha
- \frac{35495}{48} \alpha^3
- \frac{13651}{27} \Nf^2
- \frac{5535}{4} \zeta_5 \alpha
\right. \nonumber \\
&& \left. ~~~~
- \frac{3200}{9} \zeta_5 \Nf
- \frac{2385}{8} \zeta_5 \alpha^2
- \frac{1805}{12} \zeta_3 \alpha^3
- \frac{1787}{6} \Nf \alpha^2
- \frac{332}{9} \zeta_3 \Nf \alpha
\right. \nonumber \\
&& \left. ~~~~
- \frac{27}{8} \zeta_3 \alpha^4
+ \frac{3}{4} \Nf \alpha^4
+ \frac{8}{3} \Nf^3
+ \frac{51}{2} \Nf \alpha^3
+ \frac{608}{27} \zeta_3 \Nf^2 \alpha
+ \frac{749}{9} \alpha^4
\right. \nonumber \\
&& \left. ~~~~
+ \frac{1120}{27} \Nf^2 \alpha
+ \frac{1220}{27} \zeta_3 \Nf \alpha^2
+ \frac{1552}{9} \zeta_3 \Nf^2
+ \frac{1048709}{18} \alpha
+ \frac{15752321}{216} \zeta_3
\right. \nonumber \\
&& \left. ~~~~
- 17039 \zeta_3 \alpha
+ 6 \Nf^2 \alpha^2
+ 45 \zeta_5 \alpha^3
\right] a^4
\nonumber \\
&&
+ \left[
\frac{464038}{243} \Nf^3
- \frac{75504232175}{7776}
- \frac{2361380183}{1152} \zeta_3 \alpha
+ \frac{3576071485}{27648} \zeta_7
\right. \nonumber \\
&& \left. ~~~~
+ \frac{7351480225}{1728} \alpha
+ \frac{9610932889}{5832} \Nf
+ \frac{17917034005}{31104} \zeta_5
\right. \nonumber \\
&& \left. ~~~~
+ \frac{187324052147}{31104} \zeta_3
- \frac{310328447}{432} \zeta_3^2
- \frac{257106335}{324} \zeta_3 \Nf
\right. \nonumber \\
&& \left. ~~~~
- \frac{180251015}{1944} \zeta_5 \Nf
- \frac{64263505}{1152} \alpha^3
- \frac{50732065}{128} \alpha^2
- \frac{22459484}{243} \Nf^2
\right. \nonumber \\
&& \left. ~~~~
- \frac{21935425}{48} \Nf \alpha
- \frac{8970469}{3456} \zeta_3 \alpha^2
- \frac{4778536}{81} \zeta_7 \Nf
- \frac{4363519}{648} \Nf \alpha^2
\right. \nonumber \\
&& \left. ~~~~
- \frac{2413745}{64} \zeta_3^2 \alpha
- \frac{121981}{864} \alpha^5
- \frac{89389}{81} \zeta_3 \Nf^2 \alpha^2
- \frac{60928}{81} \zeta_3^2 \Nf^2
\right. \nonumber \\
&& \left. ~~~~
- \frac{35721}{256} \zeta_7 \alpha^4
- \frac{30599}{9} \zeta_3 \Nf^2 \alpha
- \frac{28096}{81} \zeta_3 \Nf^3
- \frac{26173}{8} \zeta_7 \Nf \alpha
\right. \nonumber \\
&& \left. ~~~~
- \frac{15451}{4} \zeta_3^2 \alpha^2
- \frac{12753}{64} \zeta_3^2 \alpha^3
- \frac{3015}{8} \zeta_5 \alpha^4
- \frac{2187}{64} \zeta_3^2 \alpha^4
- \frac{1600}{9} \zeta_5 \Nf^3
\right. \nonumber \\
&& \left. ~~~~
- \frac{1190}{3} \zeta_5 \Nf^2 \alpha
- \frac{641}{6} \Nf^2 \alpha^3
- \frac{452}{9} \Nf^3 \alpha
- \frac{352}{27} \Nf^4
- \frac{313}{9} \zeta_3 \Nf \alpha^4
\right. \nonumber \\
&& \left. ~~~~
- \frac{152}{3} \zeta_3 \Nf^2 \alpha^3
- \frac{135}{4} \zeta_5 \alpha^5
- \frac{128}{27} \zeta_3 \Nf^3 \alpha
- \frac{45}{4} \Nf \alpha^5
- \frac{16}{9} \zeta_3 \Nf \alpha^3
\right. \nonumber \\
&& \left. ~~~~
- \frac{9}{2} \Nf^2 \alpha^4
+ \frac{127}{2} \zeta_3^2 \Nf \alpha^2
+ \frac{483}{16} \zeta_7 \Nf \alpha^2
+ \frac{1372}{3} \zeta_7 \Nf^2
+ \frac{6169}{6} \zeta_3^2 \Nf \alpha
\right. \nonumber \\
&& \left. ~~~~
+ \frac{9973}{96} \zeta_3 \alpha^5
+ \frac{27695}{24} \zeta_5 \Nf \alpha^2
+ \frac{184009}{162} \Nf^2 \alpha^2
+ \frac{349613}{27} \Nf^2 \alpha
\right. \nonumber \\
&& \left. ~~~~
+ \frac{425405}{18} \zeta_5 \Nf \alpha
+ \frac{731917}{192} \zeta_3 \alpha^4
+ \frac{736953}{128} \zeta_7 \alpha^2
+ \frac{948548}{27} \zeta_3 \Nf^2
\right. \nonumber \\
&& \left. ~~~~
+ \frac{1161139}{144} \Nf \alpha^3
+ \frac{1449315}{1024} \zeta_7 \alpha^3
+ \frac{1489695}{128} \zeta_5 \alpha^3
+ \frac{1850845}{243} \zeta_5 \Nf^2
\right. \nonumber \\
&& \left. ~~~~
+ \frac{2410417}{288} \alpha^4
+ \frac{5255953}{36} \zeta_3 \Nf \alpha
+ \frac{6570181}{162} \zeta_3^2 \Nf
+ \frac{12827225}{384} \zeta_5 \alpha^2
\right. \nonumber \\
&& \left. ~~~~
+ \frac{15135277}{648} \zeta_3 \Nf \alpha^2
+ \frac{15328585}{384} \zeta_5 \alpha
+ \frac{51535765}{1152} \zeta_3 \alpha^3
- 90 \zeta_5 \Nf \alpha^3
\right. \nonumber \\
&& \left. ~~~~
+ \frac{54810357}{1024} \zeta_7 \alpha
- 8 \Nf^3 \alpha^2
+ 10 \zeta_5 \Nf^2 \alpha^2
+ 84 \Nf \alpha^4
\right] a^5 ~+~ O(a^6) ~.
\label{gammamqmmom5}
\end{eqnarray}
To achieve this we evaluated the Green's function where the quark mass operator
is inserted in a quark $2$-point function with (\ref{lagcf}) as the core 
Lagrangian akin to the second graph of Figure \ref{figop} where the ghost lines
are replaced by quarks. Except in the quark mass operator case the single 
momentum flow is routed through the quark legs. In this case there are $1$, 
$13$, $245$ and $5796$ Feynman graphs from one to four loops respectively. 
While the $\MSbar$ $\beta$-function and quark mass anomalous dimension are 
independent of the gauge parameter in both the linear covariant and 
Curci-Ferrari gauges, \cite{50}, this is not the case in other schemes such as 
$\mMOM$. Although we have derived these two expressions to five loops we have 
not repeated the exercise for the field and gauge parameter anomalous 
dimensions. This will only be possible when the five loop $\MSbar$ expressions 
are known in the Curci-Ferrari gauge. However we have constructed the 
conversion functions for the fields and gauge parameter and verified that the 
four loop renormalization group $\mMOM$ renormalization group functions 
determined directly and via the method that produced (\ref{betammom5}) and 
(\ref{gammamqmmom5}) are in agreement. The conversion functions to $O(a^4)$ and
the $\mMOM$ renormalization group functions for an arbitrary colour group are 
available from a data file on the arXiv page associated with this article.

One of the reasons for recording the full structure of these five loop $\mMOM$
renormalization group functions is to note that they have a particular 
property. To five loops neither $\beta^{\mMOMss}(a,\alpha)$ nor 
$\gamma_{\bar{\psi}\psi}^{\mMOMss}(a,\alpha)$ contain the numbers $\zeta_4$ or 
$\zeta_6$ for nonzero $\alpha$ unlike their $\MSbar$ counterparts. Equally the
four loop $\mMOM$ field and gauge parameter anomalous dimensions do not contain 
$\zeta_4$. This property is not dependent on the particular choice of gauge 
fixing. On the contrary the five loop $\mMOM$ renormalization group functions 
for the linear covariant gauge do not contain an even zeta when the gauge 
parameter is zero, \cite{48,58,59,60}. For $\alpha$~$\neq$~$0$ $\zeta_4$ and 
$\zeta_6$ are present to five loops in the linear covariant gauge case unlike 
the Curci-Ferrari gauge. The absence of even zetas has been the subject of 
intense interest in recent years but is not limited to the $\mMOM$ scheme or
gauge theories. See, for example, \cite{68,69,70,71}. Indeed the same feature 
occurs in the $\MOMts$ schemes introduced in \cite{72} and discussed in
\cite{69,70,71,73}. Those $\MOMts$ schemes share the same property in that at 
the subtraction point after renormalization there are no $O(a)$ corrections. 
The absence of even zetas for the Curci-Ferrari gauge to five loops is not a 
check in itself but it does add to our overview on the structure of schemes in 
different gauges and indicates the feature is not tied to a specific covariant 
gauge fixing in the case of gauge theories.

\sect{Beyond five loops.}

While the main focus to this point has been the establishment of the core
Curci-Ferrari gauge renormalization group functions, particular properties of
(\ref{lagcf}) potentially open up an avenue to access the six loop $\MSbar$ QCD
$\beta$-function. We devote this section to a summary of the process and weigh
up several strategies that could be pursued recording pros and cons for each. 
In doing so we will indicate one route that might be the most efficient. Such a
preliminary analysis is worth considering since to actually carry out a six 
loop computation requires planning as well as computer resources. Although the 
latter may not be immediately available the development of Feynman graph 
integration tools in recent years, such as {\sc Forcer} \cite{44,45}
and the Laporta reduction method \cite{74}, would assist and inevitably be 
improved over time. Indeed the Laporta approach has been significantly 
streamlined in recent years by incorporating ideas from pure mathematics. To 
appreciate the strategy we have in mind it is worth recalling some of the 
approaches to compute the QCD $\beta$-function at low loop order but beyond the
results of \cite{75,76,77,78}. In particular the three loop $\beta$-function
determined in \cite{79} illustrates some of the key issues to be accommodated 
when faced with, for that period, limited resources. The ghost and gluon 
$2$-point functions as well as the ghost-gluon vertex were renormalized in the 
Feynman gauge in \cite{79}. By this we mean the canonical linear covariant 
gauge where $\alpha$~$=$~$1$. As is evident from (\ref{cfpropsm}) the gluon 
propagator reduces to a simple scalar field propagator in both the massless and
massive case. Moreover it reduces the number of integrations by parts that have
to be effected in a Feynman integral from the longitudinal term of 
(\ref{cfpropsm}) if $\alpha$ was not unity. An alternative strategy to extract 
the three loop $\beta$-function from these specific Green's functions would 
have been to exploit Taylor's observation \cite{37} that the ghost-gluon vertex
is finite in the Landau gauge which would only require the ghost and gluon 
field renormalizations. However that would involve the transverse projector and 
hence necessitate tedious integration by parts for the dipole component. While 
the powerful {\sc Forcer} and Laporta algorithms in effect relegate this to a 
manageable issue nowadays, at higher loop order heavy integration by parts 
would be needed. It would be more beneficial to avoid this aspect in the first 
place by exploiting the Feynman gauge simplification approach used in 
\cite{79}.

{\begin{figure}[ht]
\begin{center}
\includegraphics[width=12.0cm,height=2.6cm]{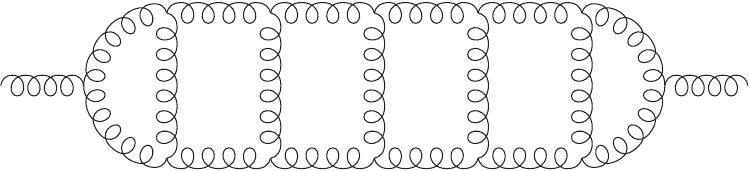}
\end{center}
\caption{Six loop graph contributing to gluon $2$-point function.}
\label{figaa6}
\end{figure}}

To appreciate one aspect of such a choice it is worth considering the structure 
of perhaps the most involved sub-class of graphs that will arise at six loops. 
These are the graphs which are purely gluonic one of which is illustrated in 
Figure \ref{figaa6} and are present in either the linear covariant or 
Curci-Ferrari gauges. To roughly benchmark the amount of algebra that would be 
involved one can estimate the number of terms that would contribute. For 
$\alpha$~$\neq$~$1$ each gluon propagator would contain two terms and lead to a
term total of $2^{17}$~$=$~$131072$. The Feynman rule for the triple gluon 
vertex involves three momenta which reduces to two using energy momentum 
conservation. Using the latter number would produce $2^{12}$~$=$~$4096$ terms 
ignoring Lorentz contractions as we are carrying out a rough estimate. For a 
linear gauge approach using Taylor's observation which is based on the Landau 
gauge the number of terms would be the virtually unmanageable total of 
$2^{29}$~$=$~$536870912$ for this gluonic graph. By contrast using the 
Curci-Ferrari gauge with $\alpha$~$=$~$1$ would reduce this by roughly five
orders of magnitude to $2^{12}$. At $L$ loops the respective numbers would be
$2^{5L-1}$ and $2^{2L}$ for the topology of Figure \ref{figaa6}.

{\begin{figure}[ht]
\begin{center}
\includegraphics[width=12.0cm,height=3.4cm]{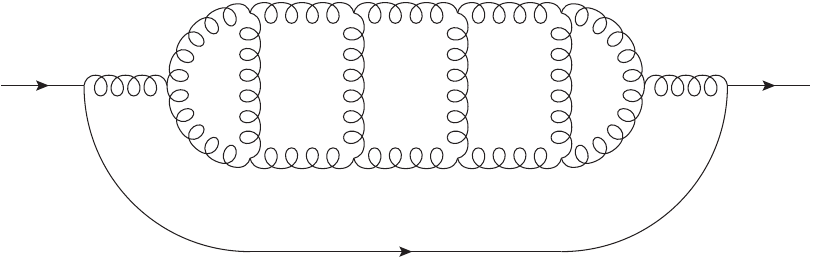}
\end{center}
\caption{Six loop graph contributing to ghost $2$-point function.}
\label{figcc6}
\end{figure}}

For balance it is appropriate to examine a similar maximally gluonic graph for 
the ghost $2$-point function evaluation as the ghost interactions are not the 
same in the two gauges we are considering. One such graph is illustrated in 
Figure \ref{figcc6}. Aside from the above gluon term counting there is only one
term in the ghost propagator in both gauges. However in the linear covariant 
gauge a ghost-gluon vertex has one term in contrast to the two of the 
Curci-Ferrari gauge ghost-gluon vertex. In the linear gauge the graph of Figure
\ref{figcc6} would have $2^{16}$~$=$~$65536$ terms for $\alpha$~$\neq$~$1$ 
together with $2^{10}$~$=$~$1024$ terms from the triple gluon vertices. While 
the total of $2^{26}$~$=$~$67108864$ is an order of magnitude less than the 
previous case it is still substantial. By contrast in the Curci-Ferrari gauge 
with $\alpha$~$=$~$1$ the total of terms would again be the same as the purely 
gluonic graph at $2^{12}$ by this rough benchmarking method. It is rough in the
sense that it is not possible to take account of simplifications that would 
follow by using momentum conservation at the internal vertices for example of 
these two graphs or for other topologies. The respective numbers at $L$ loops
for the topology of Figure \ref{figcc6} are $2^{5L-4}$ and $2^{2L}$. One final 
comment on this term counting benchmark is to note the number of purely gluonic
graphs that will arise at each loop order which is provided in Table 
\ref{aa2ptgluon}. The total at each loop order includes graphs with quartic 
gluon interactions as well as graphs with self-energy subgraphs. The graphs 
comprising these two classes would not have the same level of term number swell
from the momenta in the vertex Feynman rules as those of Figures \ref{figaa6} 
and \ref{figcc6}. Of the six loop total in Table \ref{aa2ptgluon} there are 
$856$ graphs with only $3$-point vertices and no self-energy corrections.

{\begin{table}[hb]
\begin{center}
\begin{tabular}{|c||r|r|r|r|r|r|}
\hline
\multicolumn{1}{|c||}{$L$} & 
\multicolumn{1}{|c|}{$1$} & 
\multicolumn{1}{|c|}{$2$} & 
\multicolumn{1}{|c|}{$3$} & 
\multicolumn{1}{|c|}{$4$} & 
\multicolumn{1}{|c|}{$5$} & 
\multicolumn{1}{|c|}{$6$} \\
\hline
\rule{0pt}{12pt}
$\left. \langle A^a_\mu A^b_\nu \rangle \right|_{g}$ & 
$1$ & $6$ & $49$ & $593$ & $8974$ & $161602$ \\
\hline
\end{tabular}
\end{center}
\begin{center}
\caption{Number of graphs contributing to the gluon $2$-point function that are
purely gluonic.}
\label{aa2ptgluon}
\end{center}
\end{table}}

The Curci-Ferrari gauge discussed here appears to offer a viable route to the
six loop QCD $\beta$-function that would substantially reduce the amount of 
heavy integration that would be needed. The key is the Slavnov-Taylor identity 
(\ref{zgccg}). While it contains three independent renormalization constants, 
$\{ Z_A, Z_\alpha, Z_c \}$, these are determined from only two $2$-point 
functions. Though that of the gluon involves two projections to isolate the 
transverse and longitudinal components which are related to the first two 
elements of the set. While not having to evaluate any $3$-point functions a 
substantial amount of integration would have to be carried out for the gluon 
$2$-point function. The choice of $\alpha$~$=$~$1$ offered in the Curci-Ferrari
gauge would reduce this. On the other hand the number of Feynman graphs that 
would need to be evaluated to deduce $Z_A$, $Z_\alpha$ and $Z_c$ would be 
larger than that of the linear covariant gauge due to the quartic ghost 
interaction. However in effect this is a scalar field albeit Grassmann which 
would have a minimal effect on the integration to be carried out as the 
interaction does not involve loop momenta. To appreciate the size of a six loop
renormalization it is customary to determine the number of Feynman graphs that 
would have to be evaluated first. In Table \ref{cf56qg} we have provided this
analysis with the breakdown of graph numbers contributing to the relevant 
Green's functions for the two separate gauges for the six loop $\beta$-function
as well as the respective overall totals. For the linear covariant gauge the 
ghost-gluon vertex has been included if the $\alpha$~$=$~$1$ strategy is used.
If the Taylor theorem prescription is followed at six loops for the Landau
gauge then the total number of graphs is $4232678$ which is substantially less 
than for the $\alpha$~$=$~$1$ gauge but would involve a large amount of 
integration because of the dipole term in the gluon propagator. While the total
number of graphs for a Landau gauge calculation is one third less than using 
the Curci-Ferrari gauge that gauge has the advantage of no dipole propagators 
and thus less integration. Of course this graph counting analysis also does not
take into account the finer detail of streamlining an actual evaluation using 
standard techniques such as specifying the colour group to be $SU(3)$ as was 
the case in the first five loop $\MSbar$ calculation, \cite{38}, grouping 
graphs by topologies to take advantage of cancellations and to include the 
$\epsilon$ expansion of self-energy expressions in one and two loop topologies.
The latter approach would avoid the tedious reproduction of expressions in an 
automatic Feynman integration set up that can be deduced by elementary means. 
The grouping of graphs at six loops would also be beneficial particularly for 
topologies which include a quartic ghost interaction. This is because the
colour factors would match those of the related quartic gluon ones. To close 
this aspect of the discussion we note that we reproduced the original four loop
$\MSbar$ QCD $\beta$-function of \cite{42,43} using (\ref{lagcf}) and
{\sc Forcer} but with $\alpha$~$=$~$1$. By this we mean that the four loop 
graphs of the two $2$-point functions were determined for this value and $Z_g$ 
deduced at four loops by using the $\alpha$~$\neq$~$1$ lower loop graphs 
without the need to calculate any vertex function.

{\begin{table}[ht]
\begin{center}
\begin{tabular}{|c||r|r|r|r||r|r|r|}
\hline
& \multicolumn{4}{|c||}{Linear gauge} & 
\multicolumn{3}{|c|}{Curci Ferrari gauge} \\
\hline
\rule{0pt}{12pt}
\!$L$ & $\langle A^a_\mu A^b_\nu \rangle$ & 
$\langle c^a \bar{c}^b \rangle$ & $\langle A^a_\mu \bar{c}^b c^c \rangle$ & 
Total & $\langle A^a_\mu A^b_\nu \rangle$ & $\langle c^a \bar{c}^b \rangle$ & 
Total \\
\hline
$5$ & $120074$ & $32209$ & $500559$ & $652842$ & $145870$ & $60323$ & 
$206193$ \\
$6$ & $3375438$ & $857240$ & $16215639$ & $19752679$ & $4298084$ & 
$1739411$ & $6037495$ \\
\hline
\end{tabular}
\end{center}
\begin{center}
\caption{Comparison of the number of graphs at five and six loops for the gluon
and ghost $2$-point functions and their vertex in the linear covariant gauge 
and for the gluon and ghost $2$-point functions in the Curci-Ferrari gauge plus
the separate graph total contributing to the $\beta$-function evaluation in
each gauge.}
\label{cf56qg}
\end{center}
\end{table}}

It is worth mentioning one other potential advantage of using (\ref{lagcf})
which rests in the technique used to actually evaluate the Feynman graphs. In 
the five loop $\beta$-function computations of \cite{38,39,41} the fields were 
treated as massless. When coupled to the automatic infrared rearrangement 
technique \cite{80,81,82,83}, the {\sc Forcer} package was powerful enough to 
determine the $\epsilon$ expansion of the necessary $2$-point functions and 
relevant vertex functions to the requiste orders to deduce the $\beta$-function
to five loops in the $\MSbar$ scheme. At the next order it is not clear whether
{\sc Forcer} would be sufficient to extend that five loop approach or if a new
massless integration routine would need to be developed. In either case a set
of massless, and possibly primitive, master intgrals would need to be 
determined to the simple pole in $\epsilon$. While a differential equation
approach coupled with the use of PSLQ, \cite{84}, is now a standard method 
currently for such master evaluation, other techniques might offer a better 
strategy. For instance the $\beta$-function of $\phi^4$ theory is available to 
{\em seven} loops, \cite{85,86}. Indeed the seven loop work proceeded entirely 
with the use of graphical functions \cite{86,87,88} and demonstated that 
multiple zetas arise at these orders. More crucially within these works it is 
inevitable that a substantial number of six loop massless master integrals are 
already implicitly available.

An alternative approach to evaluate the QCD renormalization group functions at
five loops was used in \cite{40,67,89}. The initial step of the process was to 
introduce a spurious mass into the propagators in a systematic way motivated by
the strategy provided in \cite{90,91}. The same mass was used for all the 
fields reducing the computation to single mass scale vacuum bubble graphs. 
These were reduced to a basis of five loop master integrals via the Laporta 
algorithm \cite{74}. The masters were determined to the requisite order in 
$\epsilon$ via high precision numerical evaluation and matching to a number 
basis using PSLQ. While this is straightforward for a nongauge theory, it was 
recognized in \cite{91} that such a spurious mass could invalidate gauge 
symmetry. However this is circumvented by allowing for the renormalization of 
this extra mass itself. By contrast in the Curci-Ferrari model a BRST invariant
mass can be included at the Lagrangian level at the outset, \cite{16}. It 
naturally avoids nonrenormalizability difficulties of a nonabelian gauge 
field having a mass without spontaneous symmetry breaking. As discussed earlier
the transverse and longitudinal components of the gluon acquire a mass, 
(\ref{cfpropsm}), with that of the latter component matched by the ghost but 
different from the gluon. The disadvantage of using the approach of 
\cite{90,91} in that instance is that the vacuum bubble expansion would require
the evaluation of vaccuum bubbles with {\em two} scales on top of the extra 
work needed to effect the Laporta reduction in the first place. However for a 
six loop evaluation with $\alpha$~$=$~$1$ we note the explicit obvious 
propagator simplification is
\begin{eqnarray}
\left. \langle A^a_\mu(p) A^b_\nu(-p) \rangle \right|_{\alpha=1}
&=& -~ \frac{\eta_{\mu\nu} \delta^{ab}}{(p^2+m^2)} \nonumber \\
\left. \langle c^a(p) \bar{c}^b(-p) \rangle \right|_{\alpha=1} 
&=& \frac{\delta^{ab}}{(p^2+m^2)} ~.
\end{eqnarray}
Moreover in the Curci-Ferrari gauge the actual renormalization of $m$ is 
already known by the Slavnov-Taylor identity of (\ref{zop}). It is determined 
by the gluon and ghost field renormalizations. So the treatment of the spurious
mass in the massive bubble expansion would be more naturally controlled than 
for the linear covariant gauge case. While these remarks are clearly apt for a 
Yang-Mills computation the inclusion of a quark with the same mass as the gluon
ought not to be a problem since proceeding in this way its treatment would fall
into the same class used in \cite{40,67,89}. On the separate issue of deducing 
the anomalous dimension of the actual quark mass itself or equivalently the 
quark mass operator, the Curci-Ferrari gauge would not offer a significant 
advantage. This is because the operator anomalous dimension is independent of 
the gauge parameter in the $\MSbar$ scheme and therefore can be deduced by 
using $\alpha$~$=$~$1$ in either the linear covariant or Curci-Ferrari gauges. 
Given the way the high order evaluation of the QCD renormalization group 
functions have proceeded with the determination of the $\MSbar$ quark mass 
anomalous dimension preceding the $\beta$-function, \cite{64,65,66,67}, the 
Curci-Ferrari gauge could instead be used as a testing ground for any six loop 
coded algorithm with $\alpha$~$=$~$1$. However in the linear gauge there would 
be fewer Feynman graphs to evaluate for this anomalous dimension. Therefore the 
advantage of choosing the Curci-Ferrari gauge over that of the linear one for 
high order computations may be more beneficial for situations where there are 
gauge independent quantities to determine which are predominantly gluonic. One 
such example is the anomalous dimensions of the twist-$2$ flavour singlet 
gluonic Wilson operators used in deep inelastic scattering.

\sect{Discussion.}

There are two main themes running through this article. The first is the 
renormalization of QCD in the Curci-Ferrari gauge, \cite{16}, to a new order 
which is four loops. Although it is a nonlinear covariant gauge with an extra 
interaction beyond that of the canonical linear covariant gauge, it has several
interesting properties from the point of view of renormalization. For instance 
the gauge admits a BRST invariant dimension two gluon and ghost mass operator 
and the renormalization of this operator is not independent of that of the 
ghost and gluon fields as well as the gauge parameter, \cite{29,30,31,33,34}. 
Indeed there are two particularly useful Slavnov-Taylor identities \cite{29,30}
one of which governs this mass operator. The other is a generalization of 
Taylor's theorem concerning the nonrenormalization of the ghost-gluon vertex in
the Landau gauge. As part of our explicit four loop renormalization we have 
checked that both identities are satisfied in the $\MSbar$ scheme for all 
values of the Curci-Ferrari gauge parameter. As a corollary to the vertex 
identity we generalized the $\mMOM$ scheme of \cite{48} in a linear covariant 
gauge to the Curci-Ferrari gauge. One consequence of the four loop 
renormalization was that we were able to deduce the {\em five} loop 
$\beta$-function and quark mass anomalous dimension in the $\mMOM$ scheme which
will be used to update the analysis of \cite{51}.

The second theme of the article was more forward looking as we examined the
potential that the Curci-Ferrari gauge offers for any future attempt to extract
the six loop $\MSbar$ QCD $\beta$-function. The {\it b\^{e}te-noire} of any 
high order multiloop automatic computation is the amount of time required to 
handle the algebra associated with the gluon propagator due to its dipole 
component. This can be circumvented by choosing the Feynman gauge for the 
canonical gauge fixing but as was evident in \cite{79} at three loops in order 
to extract the coupling renormalization constant the ghost-gluon vertex needs 
to be evaluated. The Curci-Ferrari gauge and in particular the vertex 
Slavnov-Taylor identity avoids having to determine the ghost-gluon vertex at 
six loops at all as well as excluding the need to include the dipole term of 
the gluon propagator by choosing $\alpha$~$=$~$1$. While the number of Feynman 
graphs would still be large at around four million it is almost five times less
than having to compute the vertex for the same value of $\alpha$ in the linear 
gauge. Any attempt to tackle such a mammoth renormalization may have to wait 
for a powerful enough algorithm and computing hardware to become available. For
practical purposes the two techniques that were employed to determine the five 
loop $\MSbar$ $\beta$-function should serve as a launch direction. Finally we 
note that while such a new order coupling constant renormalization would be a 
long term commitment, the Curci-Ferrari gauge could be equally beneficial for 
current high order computations. For instance there has been activity in 
calculating low moments of flavour nonsinglet gauge invariant twist-$2$ Wilson 
operators used in deep inelastic scattering, \cite{92}. The next stage would be
to construct the low moment anomalous dimensions for the flavour singlet 
operators and specifically the gluonic ones. As these latter operators are 
structurally parallel to the field strength term of (\ref{lagcf}) the heavy 
integration by parts associated with the gluon propagator dipole term would 
naturally be avoided in the Curci-Ferrari gauge when $\alpha$~$=$~$1$. In
employing such an approach it should be feasible to expand the number of
moments beyond the few known low ones for gluonic operators not only at four 
loops but at five loops as well.

\vspace{1cm}
\noindent
{\bf Data Availability Statement.} The data representing the main results here 
are accessible in electronic form from the arXiv ancillary directory associated
with the article.

\vspace{0.5cm}
\noindent
{\bf Acknowledgements.} The author thanks A. Maier, R. Mason, P. Marquard and
N. Wschebor for useful discussions, the DESY Zeuthen theory group for their 
hospitality where part of this work was initiated and Kolleg Mathematik Physik 
Berlin for a Visiting Scholarship. The research was carried out with the 
support of the STFC Consolidated Grant ST/X000699/1 and was undertaken on 
{\sc Barkla}, part of the High Performance Computing facilities at the 
University of Liverpool, UK. For the purpose of open access, the author has 
applied a Creative Commons Attribution (CC-BY) licence to any Author Accepted 
Manuscript version arising. 

\appendix

\sect{$\MSbar$ scheme renormalization group functions.}

In this Appendix we record the four loop $\MSbar$ renormalization group 
functions for an arbitrary colour group. The gauge parameter dependence is 
included since the $\alpha$~$=$~$0$ expressions reproduce the known Landau 
gauge ones. We have
\begin{eqnarray}
\gamma_A(a,\alpha) &=&
\left[
\frac{4}{3} T_F \Nf
- \frac{13}{6} C_A
+ \frac{1}{2} C_A \alpha
\right] a
\nonumber \\
&&
+ \left[
5 C_A T_F \Nf
- \frac{59}{8} C_A^2
+ \frac{1}{8} C_A^2 \alpha^2
+ \frac{11}{8} C_A^2 \alpha
+ 4 C_F T_F \Nf
\right] a^2
\nonumber \\
&&
+ \left[
24 \zeta_3 C_A C_F T_F \Nf
- \frac{9965}{288} C_A^3
- \frac{76}{9} C_A T_F^2 \Nf^2
- \frac{44}{9} C_F T_F^2 \Nf^2
+ \frac{3}{4} \zeta_3 C_A^3 \alpha
+ \frac{3}{64} C_A^3 \alpha^3
\right. \nonumber \\
&& \left. ~~~~
+ \frac{5}{18} C_A C_F T_F \Nf
+ \frac{9}{16} \zeta_3 C_A^3
+ \frac{101}{128} C_A^3 \alpha^2
+ \frac{167}{32} C_A^3 \alpha
+ \frac{911}{18} C_A^2 T_F \Nf
\right. \nonumber \\
&& \left. ~~~~
- 18 \zeta_3 C_A^2 T_F \Nf
- 2 C_F^2 T_F \Nf
- 2 C_A^2 T_F \Nf \alpha
\right] a^3
\nonumber \\
&&
+ \left[
120 \zeta_5 C_A^2 C_F T_F \Nf
- \frac{10655437}{62208} C_A^4
- \frac{146831}{972} C_A^2 C_F T_F \Nf
- \frac{47665}{512} \zeta_5 C_A^4
\right. \nonumber \\
&& \left. ~~~~
- \frac{15082}{243} C_A C_F T_F^2 \Nf^2
- \frac{10235}{768} \zeta_5 C_A^4 \alpha
- \frac{10185}{64} \zeta_5 \frac{d_A^{abcd} d_A^{abcd}}{\NA}
- \frac{6674}{81} C_A^2 T_F^2 \Nf^2
\right. \nonumber \\
&& \left. ~~~~
- \frac{3563}{12} \zeta_3 C_A^3 T_F \Nf
- \frac{1556}{243} C_A^3 T_F \Nf \alpha
- \frac{1420}{243} C_A T_F^3 \Nf^3
- \frac{1352}{27} C_F^2 T_F^2 \Nf^2
\right. \nonumber \\
&& \left. ~~~~
- \frac{1232}{243} C_F T_F^3 \Nf^3
- \frac{1229}{486} C_A^2 T_F^2 \Nf^2 \alpha
- \frac{1168}{9} \zeta_3 C_A C_F T_F^2 \Nf^2
- \frac{989}{12} \zeta_3 \frac{d_A^{abcd} d_A^{abcd}}{\NA}
\right. \nonumber \\
&& \left. ~~~~
- \frac{863}{24} C_A^2 C_F T_F \Nf \alpha
- \frac{512}{3} \zeta_3 \Nf^2 \frac{d_F^{abcd} d_F^{abcd}}{\NA}
- \frac{512}{9} \Nf \frac{d_F^{abcd} d_A^{abcd}}{\NA}
\right. \nonumber \\
&& \left. ~~~~
- \frac{475}{32} \zeta_5 \frac{d_A^{abcd} d_A^{abcd}}{\NA} \alpha
- \frac{203}{64} \zeta_4 C_A^4 \alpha
- \frac{115}{144} C_A^3 T_F \Nf \alpha^2
- \frac{99}{32} \zeta_4 C_A^4
- \frac{7}{2} \zeta_4 C_A^3 T_F \Nf \alpha
\right. \nonumber \\
&& \left. ~~~~
- \frac{5}{24} \zeta_3 C_A^3 T_F \Nf \alpha^2
- \frac{5}{96} \zeta_3 C_A^4 \alpha^3
- \frac{3}{4} \frac{d_A^{abcd} d_A^{abcd}}{\NA} \alpha
- \frac{3}{16} \frac{d_A^{abcd} d_A^{abcd}}{\NA} \alpha^2
- \frac{1}{768} \zeta_3 C_A^4 \alpha^4
\right. \nonumber \\
&& \left. ~~~~
+ \frac{1}{8} \zeta_3 \frac{d_A^{abcd} d_A^{abcd}}{\NA} \alpha^3
+ \frac{1}{32} \zeta_3 \frac{d_A^{abcd} d_A^{abcd}}{\NA} \alpha^2
+ \frac{1}{32} \zeta_3 \frac{d_A^{abcd} d_A^{abcd}}{\NA} \alpha^4
+ \frac{3}{32} \zeta_4 C_A^4 \alpha^2
\right. \nonumber \\
&& \left. ~~~~
+ \frac{5}{6} \zeta_5 C_A^4 \alpha^2
+ \frac{5}{8} \zeta_5 \frac{d_A^{abcd} d_A^{abcd}}{\NA} \alpha^3
+ \frac{5}{96} \zeta_5 C_A^4 \alpha^3
+ \frac{8}{3} \zeta_3 C_A^2 T_F^2 \Nf^2 \alpha
\right. \nonumber \\
&& \left. ~~~~
+ \frac{19}{768} C_A^4 \alpha^4
+ \frac{64}{9} \zeta_3 C_A T_F^3 \Nf^3
+ \frac{169}{768} \zeta_3 C_A^4 \alpha^2
+ \frac{185}{32} \zeta_5 \frac{d_A^{abcd} d_A^{abcd}}{\NA} \alpha^2
\right. \nonumber \\
&& \left. ~~~~
+ \frac{316}{9} \zeta_3 C_A^2 T_F^2 \Nf^2
+ \frac{355}{768} C_A^4 \alpha^3
+ \frac{449}{16} \zeta_3 \frac{d_A^{abcd} d_A^{abcd}}{\NA} \alpha
+ \frac{659}{144} \frac{d_A^{abcd} d_A^{abcd}}{\NA}
\right. \nonumber \\
&& \left. ~~~~
+ \frac{704}{9} \Nf^2 \frac{d_F^{abcd} d_F^{abcd}}{\NA}
+ \frac{704}{9} \zeta_3 C_F^2 T_F^2 \Nf^2
+ \frac{801}{8} \zeta_4 C_A^3 T_F \Nf
+ \frac{980}{9} \zeta_3 C_A C_F^2 T_F \Nf
\right. \nonumber \\
&& \left. ~~~~
+ \frac{1155}{64} \zeta_3 C_A^4 \alpha
+ \frac{1376}{3} \zeta_3 \Nf \frac{d_F^{abcd} d_A^{abcd}}{\NA}
+ \frac{2240}{9} \zeta_3 C_A^2 C_F T_F \Nf
+ \frac{8765}{2304} C_A^4 \alpha^2
\right. \nonumber \\
&& \left. ~~~~
+ \frac{10847}{54} C_A C_F^2 T_F \Nf
+ \frac{50669}{768} \zeta_3 C_A^4
+ \frac{318907}{864} C_A^3 T_F \Nf
+ \frac{2184485}{62208} C_A^4 \alpha
\right. \nonumber \\
&& \left. ~~~~
- 240 \zeta_5 C_A C_F^2 T_F \Nf
- 132 \zeta_4 C_A^2 C_F T_F \Nf
- 46 C_F^3 T_F \Nf
- 36 \zeta_4 C_A^2 T_F^2 \Nf^2
\right. \nonumber \\
&& \left. ~~~~
- 24 \zeta_3 C_A^3 T_F \Nf \alpha
+ 6 \zeta_4 C_A^2 C_F T_F \Nf \alpha
+ 28 \zeta_3 C_A^2 C_F T_F \Nf \alpha
+ 48 \zeta_4 C_A C_F T_F^2 \Nf^2
\right. \nonumber \\
&& \left. ~~~~
+ 110 \zeta_5 C_A^3 T_F \Nf
+ 120 \zeta_5 \Nf \frac{d_F^{abcd} d_A^{abcd}}{\NA}
\right] a^4 ~+~ O(a^5) \nonumber \\
\gamma_\alpha(a,\alpha) &=&
\left[
\frac{13}{6} C_A
- \frac{4}{3} T_F \Nf
- \frac{1}{4} C_A \alpha
\right] a
\nonumber \\
&&
+ \left[
\frac{59}{8} C_A^2
- \frac{17}{16} C_A^2 \alpha
- \frac{1}{16} C_A^2 \alpha^2
- 5 C_A T_F \Nf
- 4 C_F T_F \Nf
\right] a^2
\nonumber \\
&&
+ \left[
2 C_F^2 T_F \Nf
- \frac{911}{18} C_A^2 T_F \Nf
- \frac{467}{128} C_A^3 \alpha
- \frac{31}{64} C_A^3 \alpha^2
- \frac{9}{16} \zeta_3 C_A^3
- \frac{5}{18} C_A C_F T_F \Nf
\right. \nonumber \\
&& \left. ~~~~
- \frac{3}{4} \zeta_3 C_A^3 \alpha
- \frac{3}{128} C_A^3 \alpha^3
+ \frac{17}{16} C_A^2 T_F \Nf \alpha
+ \frac{44}{9} C_F T_F^2 \Nf^2
+ \frac{76}{9} C_A T_F^2 \Nf^2
\right. \nonumber \\
&& \left. ~~~~
+ \frac{9965}{288} C_A^3
- 24 \zeta_3 C_A C_F T_F \Nf
+ 18 \zeta_3 C_A^2 T_F \Nf
\right] a^3
\nonumber \\
&&
+ \left[
240 \zeta_5 C_A C_F^2 T_F \Nf
- \frac{318907}{864} C_A^3 T_F \Nf
- \frac{50669}{768} \zeta_3 C_A^4
- \frac{24199}{972} C_A^4 \alpha
\right. \nonumber \\
&& \left. ~~~~
- \frac{10847}{54} C_A C_F^2 T_F \Nf
- \frac{5891}{384} \zeta_3 C_A^4 \alpha
- \frac{2741}{1152} C_A^4 \alpha^2
- \frac{2240}{9} \zeta_3 C_A^2 C_F T_F \Nf
\right. \nonumber \\
&& \left. ~~~~
- \frac{1376}{3} \zeta_3 \Nf \frac{d_F^{abcd} d_A^{abcd}}{\NA}
- \frac{980}{9} \zeta_3 C_A C_F^2 T_F \Nf
- \frac{801}{8} \zeta_4 C_A^3 T_F \Nf
- \frac{704}{9} \Nf^2 \frac{d_F^{abcd} d_F^{abcd}}{\NA}
\right. \nonumber \\
&& \left. ~~~~
- \frac{704}{9} \zeta_3 C_F^2 T_F^2 \Nf^2
- \frac{659}{144} \frac{d_A^{abcd} d_A^{abcd}}{\NA}
- \frac{439}{1536} C_A^4 \alpha^3
- \frac{316}{9} \zeta_3 C_A^2 T_F^2 \Nf^2
\right. \nonumber \\
&& \left. ~~~~
- \frac{169}{8} \zeta_3 \frac{d_A^{abcd} d_A^{abcd}}{\NA} \alpha
- \frac{64}{9} \zeta_3 C_A T_F^3 \Nf^3
- \frac{55}{48} \zeta_5 C_A^4 \alpha^2
- \frac{25}{4} \zeta_5 \frac{d_A^{abcd} d_A^{abcd}}{\NA} \alpha^2
\right. \nonumber \\
&& \left. ~~~~
- \frac{19}{1536} C_A^4 \alpha^4
- \frac{15}{64} \zeta_3 \frac{d_A^{abcd} d_A^{abcd}}{\NA} \alpha^3
- \frac{4}{3} \zeta_3 C_A^2 T_F^2 \Nf^2 \alpha
- \frac{1}{64} \zeta_3 \frac{d_A^{abcd} d_A^{abcd}}{\NA} \alpha^4
\right. \nonumber \\
&& \left. ~~~~
+ \frac{1}{1536} \zeta_3 C_A^4 \alpha^4
+ \frac{3}{4} \frac{d_A^{abcd} d_A^{abcd}}{\NA} \alpha
+ \frac{3}{16} \frac{d_A^{abcd} d_A^{abcd}}{\NA} \alpha^2
+ \frac{5}{4} \zeta_4 C_A^3 T_F \Nf \alpha
\right. \nonumber \\
&& \left. ~~~~
+ \frac{5}{512} \zeta_3 C_A^4 \alpha^3
+ \frac{23}{96} \zeta_3 C_A^4 \alpha^2
+ \frac{79}{288} C_A^3 T_F \Nf \alpha^2
+ \frac{99}{32} \zeta_4 C_A^4
\right. \nonumber \\
&& \left. ~~~~
+ \frac{139}{8} \zeta_3 C_A^3 T_F \Nf \alpha
+ \frac{235}{32} \zeta_5 \frac{d_A^{abcd} d_A^{abcd}}{\NA} \alpha
+ \frac{415}{128} \zeta_4 C_A^4 \alpha
+ \frac{512}{3} \zeta_3 \Nf^2 \frac{d_F^{abcd} d_F^{abcd}}{\NA}
\right. \nonumber \\
&& \left. ~~~~
+ \frac{512}{9} \Nf \frac{d_F^{abcd} d_A^{abcd}}{\NA}
+ \frac{541}{24} C_A^2 C_F T_F \Nf \alpha
+ \frac{727}{486} C_A^2 T_F^2 \Nf^2 \alpha
+ \frac{989}{12} \zeta_3 \frac{d_A^{abcd} d_A^{abcd}}{\NA}
\right. \nonumber \\
&& \left. ~~~~
+ \frac{1168}{9} \zeta_3 C_A C_F T_F^2 \Nf^2
+ \frac{1232}{243} C_F T_F^3 \Nf^3
+ \frac{1352}{27} C_F^2 T_F^2 \Nf^2
+ \frac{1420}{243} C_A T_F^3 \Nf^3
\right. \nonumber \\
&& \left. ~~~~
+ \frac{2015}{192} \zeta_5 C_A^4 \alpha
+ \frac{3563}{12} \zeta_3 C_A^3 T_F \Nf
+ \frac{6674}{81} C_A^2 T_F^2 \Nf^2
+ \frac{10185}{64} \zeta_5 \frac{d_A^{abcd} d_A^{abcd}}{\NA}
\right. \nonumber \\
&& \left. ~~~~
+ \frac{15082}{243} C_A C_F T_F^2 \Nf^2
+ \frac{41537}{7776} C_A^3 T_F \Nf \alpha
+ \frac{47665}{512} \zeta_5 C_A^4
+ \frac{146831}{972} C_A^2 C_F T_F \Nf
\right. \nonumber \\
&& \left. ~~~~
+ \frac{10655437}{62208} C_A^4
- 120 \zeta_5 \Nf \frac{d_F^{abcd} d_A^{abcd}}{\NA}
- 120 \zeta_5 C_A^2 C_F T_F \Nf
- 110 \zeta_5 C_A^3 T_F \Nf
\right. \nonumber \\
&& \left. ~~~~
- 48 \zeta_4 C_A C_F T_F^2 \Nf^2
- 20 \zeta_3 C_A^2 C_F T_F \Nf \alpha
- 3 \zeta_4 C_A^2 C_F T_F \Nf \alpha
+ \zeta_3 \frac{d_A^{abcd} d_A^{abcd}}{\NA} \alpha^2
\right. \nonumber \\
&& \left. ~~~~
+ 36 \zeta_4 C_A^2 T_F^2 \Nf^2
+ 46 C_F^3 T_F \Nf
+ 132 \zeta_4 C_A^2 C_F T_F \Nf
\right] a^4 ~+~ O(a^5) \nonumber \\
\gamma_c(a,\alpha) &=&
\left[
\frac{1}{4} C_A \alpha
- \frac{3}{4} C_A
\right] a
+ \left[
\frac{5}{6} C_A T_F \Nf
- \frac{95}{48} C_A^2
- \frac{1}{16} C_A^2 \alpha
+ \frac{1}{16} C_A^2 \alpha^2
\right] a^2
\nonumber \\
&&
+ \left[
9 \zeta_3 C_A^2 T_F \Nf
- \frac{15817}{1728} C_A^3
- \frac{9}{32} \zeta_3 C_A^3
- \frac{7}{8} C_A^2 T_F \Nf \alpha
- \frac{3}{8} \zeta_3 C_A^3 \alpha
+ \frac{3}{128} C_A^3 \alpha^3
\right. \nonumber \\
&& \left. ~~~~
+ \frac{17}{32} C_A^3 \alpha
+ \frac{35}{27} C_A T_F^2 \Nf^2
+ \frac{45}{4} C_A C_F T_F \Nf
+ \frac{55}{256} C_A^3 \alpha^2
+ \frac{97}{108} C_A^2 T_F \Nf
\right. \nonumber \\
&& \left. ~~~~
- 12 \zeta_3 C_A C_F T_F \Nf
\right] a^3
\nonumber \\
&&
+ \left[
120 \zeta_5 C_A C_F^2 T_F \Nf
- \frac{319561}{4608} C_A^4
- \frac{140743}{4608} \zeta_3 C_A^4
- \frac{1387}{384} \zeta_3 C_A^4 \alpha
- \frac{801}{16} \zeta_4 C_A^3 T_F \Nf
\right. \nonumber \\
&& \left. ~~~~
- \frac{779}{972} C_A^2 T_F^2 \Nf^2 \alpha
- \frac{628}{81} C_A^2 T_F^2 \Nf^2
- \frac{609}{8} \zeta_3 \frac{d_A^{abcd} d_A^{abcd}}{\NA}
- \frac{485}{64} \zeta_5 \frac{d_A^{abcd} d_A^{abcd}}{\NA} \alpha
\right. \nonumber \\
&& \left. ~~~~
- \frac{425}{48} C_A^2 C_F T_F \Nf \alpha
- \frac{271}{12} C_A C_F^2 T_F \Nf
- \frac{245}{64} \zeta_5 \frac{d_A^{abcd} d_A^{abcd}}{\NA} \alpha^2
- \frac{187}{288} C_A^3 T_F \Nf \alpha^2
\right. \nonumber \\
&& \left. ~~~~
- \frac{115}{27} C_A C_F T_F^2 \Nf^2
- \frac{32}{9} \zeta_3 C_A T_F^3 \Nf^3
- \frac{25}{24} \zeta_5 C_A^4 \alpha^2
- \frac{15}{256} \zeta_3 C_A^4 \alpha^3
\right. \nonumber \\
&& \left. ~~~~
- \frac{11}{4} \zeta_4 C_A^3 T_F \Nf \alpha
- \frac{9}{32} \zeta_3 \frac{d_A^{abcd} d_A^{abcd}}{\NA} \alpha^3
- \frac{5}{4} \zeta_3 C_A^3 T_F \Nf \alpha
- \frac{5}{16} \zeta_3 C_A^3 T_F \Nf \alpha^2
\right. \nonumber \\
&& \left. ~~~~
- \frac{5}{32} \zeta_3 \frac{d_A^{abcd} d_A^{abcd}}{\NA} \alpha
- \frac{1}{1536} \zeta_3 C_A^4 \alpha^4
+ \frac{1}{64} \zeta_3 \frac{d_A^{abcd} d_A^{abcd}}{\NA} \alpha^4
+ \frac{3}{8} \frac{d_A^{abcd} d_A^{abcd}}{\NA} \alpha
\right. \nonumber \\
&& \left. ~~~~
+ \frac{3}{32} \frac{d_A^{abcd} d_A^{abcd}}{\NA} \alpha^2
+ \frac{4}{3} \zeta_3 C_A^2 T_F^2 \Nf^2 \alpha
+ \frac{5}{64} \zeta_5 C_A^4 \alpha^3
+ \frac{9}{64} \zeta_4 C_A^4 \alpha^2
\right. \nonumber \\
&& \left. ~~~~
+ \frac{15}{16} \zeta_5 \frac{d_A^{abcd} d_A^{abcd}}{\NA} \alpha^3
+ \frac{19}{1536} C_A^4 \alpha^4
+ \frac{69}{32} \frac{d_A^{abcd} d_A^{abcd}}{\NA}
+ \frac{99}{64} \zeta_4 C_A^4
\right. \nonumber \\
&& \left. ~~~~
+ \frac{131}{64} \zeta_3 \frac{d_A^{abcd} d_A^{abcd}}{\NA} \alpha^2
+ \frac{166}{81} C_A T_F^3 \Nf^3
+ \frac{187}{1536} C_A^4 \alpha^3
+ \frac{221}{128} \zeta_4 C_A^4 \alpha
+ \frac{1243}{1536} \zeta_3 C_A^4 \alpha^2
\right. \nonumber \\
&& \left. ~~~~
+ \frac{1535}{1536} \zeta_5 C_A^4 \alpha
+ \frac{3019}{24} \zeta_3 C_A^3 T_F \Nf
+ \frac{4193}{3888} C_A^3 T_F \Nf \alpha
+ \frac{4367}{4608} C_A^4 \alpha^2
\right. \nonumber \\
&& \left. ~~~~
+ \frac{10185}{128} \zeta_5 \frac{d_A^{abcd} d_A^{abcd}}{\NA}
+ \frac{13171}{216} C_A^2 C_F T_F \Nf
+ \frac{47665}{1024} \zeta_5 C_A^4
+ \frac{295855}{5184} C_A^3 T_F \Nf
\right. \nonumber \\
&& \left. ~~~~
+ \frac{358511}{124416} C_A^4 \alpha
- 88 \zeta_3 C_A^2 C_F T_F \Nf
- 74 \zeta_3 C_A C_F^2 T_F \Nf
- 60 \zeta_5 \Nf \frac{d_F^{abcd} d_A^{abcd}}{\NA}
\right. \nonumber \\
&& \left. ~~~~
- 60 \zeta_5 C_A^2 C_F T_F \Nf
- 55 \zeta_5 C_A^3 T_F \Nf
- 30 \zeta_3 C_A^2 T_F^2 \Nf^2
- 24 \zeta_4 C_A C_F T_F^2 \Nf^2
\right. \nonumber \\
&& \left. ~~~~
+ 2 \zeta_3 C_A^2 C_F T_F \Nf \alpha
+ 3 \zeta_4 C_A^2 C_F T_F \Nf \alpha
+ 18 \zeta_4 C_A^2 T_F^2 \Nf^2
+ 40 \zeta_3 C_A C_F T_F^2 \Nf^2
\right. \nonumber \\
&& \left. ~~~~
+ 48 \zeta_3 \Nf \frac{d_F^{abcd} d_A^{abcd}}{\NA}
+ 66 \zeta_4 C_A^2 C_F T_F \Nf
\right] a^4 ~+~ O(a^5) \nonumber \\
\gamma_\psi(a,\alpha) &=&
C_F \alpha a
+ \left[
- \frac{3}{2} C_F^2
+ \frac{25}{4} C_A C_F
- 2 C_F T_F \Nf
+ 2 C_A C_F \alpha
\right] a^2
\nonumber \\
&&
+ \left[
12 \zeta_3 C_A C_F^2
- \frac{287}{9} C_A C_F T_F \Nf
- \frac{143}{4} C_A C_F^2
- \frac{69}{8} \zeta_3 C_A^2 C_F
- \frac{17}{4} C_A C_F T_F \Nf \alpha
\right. \nonumber \\
&& \left. ~~~~
+ \frac{3}{2} C_F^3
+ \frac{3}{4} \zeta_3 C_A^2 C_F \alpha
+ \frac{3}{32} C_A^2 C_F \alpha^3
+ \frac{15}{16} C_A^2 C_F \alpha^2
+ \frac{20}{9} C_F T_F^2 \Nf^2
+ \frac{263}{32} C_A^2 C_F \alpha
\right. \nonumber \\
&& \left. ~~~~
+ \frac{9155}{144} C_A^2 C_F
+ 3 C_F^2 T_F \Nf
\right] a^3
\nonumber \\
&&
+ \left[
848 \zeta_3 C_A C_F^3
- \frac{37093}{3888} C_A^2 C_F T_F \Nf \alpha
- \frac{23885}{36} C_A^2 C_F^2
- \frac{18371}{54} C_A^2 C_F T_F \Nf
\right. \nonumber \\
&& \left. ~~~~
- \frac{6209}{64} \zeta_3 C_A^3 C_F
- \frac{4145}{32} \zeta_5 C_A^3 C_F
- \frac{1076}{243} C_A C_F T_F^2 \Nf^2 \alpha
- \frac{1027}{8} C_F^4
\right. \nonumber \\
&& \left. ~~~~
- \frac{855}{8} \zeta_5 \frac{d_F^{abcd} d_A^{abcd}}{\Nc}
- \frac{767}{12} C_A C_F^2 T_F \Nf \alpha
- \frac{575}{24} \zeta_5 C_A^3 C_F \alpha
- \frac{421}{2} \zeta_3 C_A^2 C_F^2
\right. \nonumber \\
&& \left. ~~~~
- \frac{97}{32} \zeta_4 C_A^3 C_F \alpha
- \frac{69}{4} \zeta_4 C_A^2 C_F T_F \Nf
- \frac{53}{18} C_A C_F^2 T_F \Nf
- \frac{7}{2} \zeta_3 C_A^2 C_F^2 \alpha
\right. \nonumber \\
&& \left. ~~~~
- \frac{3}{2} \zeta_3 \frac{d_F^{abcd} d_A^{abcd}}{\Nc} \alpha^3
- \frac{3}{2} \zeta_3 C_A^2 C_F T_F \Nf
- \frac{1}{8} \zeta_3 \frac{d_F^{abcd} d_A^{abcd}}{\Nc} \alpha^4
+ \frac{1}{6} C_A^2 C_F T_F \Nf \alpha^2
\right. \nonumber \\
&& \left. ~~~~
+ \frac{1}{16} \zeta_3 C_A^3 C_F \alpha^3
+ \frac{1}{96} C_A^3 C_F \alpha^4
+ \frac{1}{192} \zeta_3 C_A^3 C_F \alpha^4
+ \frac{5}{24} \zeta_5 C_A^3 C_F \alpha^2
+ \frac{9}{32} \zeta_4 C_A^3 C_F \alpha^2
\right. \nonumber \\
&& \left. ~~~~
+ \frac{16}{3} \zeta_3 C_A C_F T_F^2 \Nf^2 \alpha
+ \frac{37}{24} \zeta_3 C_A^3 C_F \alpha^2
+ \frac{76}{3} C_F^3 T_F \Nf
+ \frac{97}{192} C_A^3 C_F \alpha^3
\right. \nonumber \\
&& \left. ~~~~
+ \frac{101}{2} \zeta_3 \frac{d_F^{abcd} d_A^{abcd}}{\Nc} \alpha
+ \frac{280}{81} C_F T_F^3 \Nf^3
+ \frac{293}{9} C_A C_F T_F^2 \Nf^2
+ \frac{304}{9} C_F^2 T_F^2 \Nf^2
\right. \nonumber \\
&& \left. ~~~~
+ \frac{437}{128} C_A^3 C_F \alpha^2
+ \frac{759}{16} \zeta_4 C_A^3 C_F
+ \frac{909}{32} \zeta_3 C_A^3 C_F \alpha
+ \frac{1113}{8} \zeta_3 \frac{d_F^{abcd} d_A^{abcd}}{\Nc}
\right. \nonumber \\
&& \left. ~~~~
+ \frac{5131}{12} C_A C_F^3
+ \frac{95261}{162} C_A^3 C_F
+ \frac{1596947}{31104} C_A^3 C_F \alpha
- 1440 \zeta_5 C_A C_F^3
- 400 \zeta_3 C_F^4
\right. \nonumber \\
&& \left. ~~~~
- 80 \zeta_5 C_A^2 C_F T_F \Nf
- 67 \frac{d_F^{abcd} d_A^{abcd}}{\Nc}
- 66 \zeta_4 C_A^2 C_F^2
- 64 \zeta_3 C_F^3 T_F \Nf
\right. \nonumber \\
&& \left. ~~~~
- 55 \zeta_5 \frac{d_F^{abcd} d_A^{abcd}}{\Nc} \alpha
- 37 \zeta_3 C_A^2 C_F T_F \Nf \alpha
- 32 \zeta_3 C_F^2 T_F^2 \Nf^2
- 8 \zeta_4 C_A^2 C_F T_F \Nf \alpha
\right. \nonumber \\
&& \left. ~~~~
- 5 \zeta_5 \frac{d_F^{abcd} d_A^{abcd}}{\Nc} \alpha^2
- \zeta_3 C_A^2 C_F T_F \Nf \alpha^2
+ \frac{d_F^{abcd} d_A^{abcd}}{\Nc} \alpha
+ 2 \zeta_3 \frac{d_F^{abcd} d_A^{abcd}}{\Nc} \alpha^2
\right. \nonumber \\
&& \left. ~~~~
+ 3 C_A^2 C_F^2 \alpha
+ 5 \zeta_5 C_A^2 C_F^2 \alpha
+ 12 \zeta_4 C_A C_F^2 T_F \Nf \alpha
+ 24 \zeta_4 C_A C_F^2 T_F \Nf
\right. \nonumber \\
&& \left. ~~~~
+ 32 \zeta_3 C_A C_F T_F^2 \Nf^2
+ 44 \zeta_3 C_A C_F^2 T_F \Nf \alpha
+ 128 \Nf \frac{d_F^{abcd} d_F^{abcd}}{\Nc}
+ 160 \zeta_5 C_A C_F^2 T_F \Nf
\right. \nonumber \\
&& \left. ~~~~
+ 640 \zeta_5 C_F^4
+ 785 \zeta_5 C_A^2 C_F^2
\right] a^4 ~+~ O(a^5) 
\end{eqnarray}
for the field anomalous dimensions where
\begin{equation}
d_R^{abcd} ~=~ \frac{1}{6} \mbox{Tr} \left( T^a T^{(b} T^c T^{d)}
\right)
\end{equation}
is a totally symmetric rank $4$ group tensor in the representation $R$,
\cite{93}, and appears either in the fundamental, $F$, or adjoint, $A$, 
representations. The renormalization constant of the gauge parameter is encoded
in the combination 
\begin{eqnarray}
\gamma_A(a,\alpha) ~+~ \gamma_\alpha(a,\alpha) &=&
\frac{1}{4} C_A \alpha a
+ \left[
\frac{1}{16} C_A^2 \alpha^2
+ \frac{5}{16} C_A^2 \alpha
\right] a^2
\nonumber \\
&&
+ \left[
\frac{201}{128} C_A^3 \alpha
- \frac{15}{16} C_A^2 T_F \Nf \alpha
+ \frac{3}{128} C_A^3 \alpha^3
+ \frac{39}{128} C_A^3 \alpha^2
\right] a^3
\nonumber \\
&&
+ \left[
8 \zeta_3 C_A^2 C_F T_F \Nf \alpha
- \frac{8255}{7776} C_A^3 T_F \Nf \alpha
- \frac{725}{256} \zeta_5 C_A^4 \alpha
- \frac{251}{243} C_A^2 T_F^2 \Nf^2 \alpha
\right. \nonumber \\
&& \left. ~~~~
- \frac{161}{12} C_A^2 C_F T_F \Nf \alpha
- \frac{151}{288} C_A^3 T_F \Nf \alpha^2
- \frac{65}{1536} \zeta_3 C_A^4 \alpha^3
\right. \nonumber \\
&& \left. ~~~~
- \frac{53}{8} \zeta_3 C_A^3 T_F \Nf \alpha
- \frac{15}{2} \zeta_5 \frac{d_A^{abcd} d_A^{abcd}}{\NA} \alpha
- \frac{15}{32} \zeta_5 \frac{d_A^{abcd} d_A^{abcd}}{\NA} \alpha^2
\right. \nonumber \\
&& \left. ~~~~
- \frac{9}{4} \zeta_4 C_A^3 T_F \Nf \alpha
- \frac{7}{64} \zeta_3 \frac{d_A^{abcd} d_A^{abcd}}{\NA} \alpha^3
- \frac{5}{16} \zeta_5 C_A^4 \alpha^2
\right. \nonumber \\
&& \left. ~~~~
- \frac{5}{24} \zeta_3 C_A^3 T_F \Nf \alpha^2
- \frac{1}{1536} \zeta_3 C_A^4 \alpha^4
+ \frac{1}{64} \zeta_3 \frac{d_A^{abcd} d_A^{abcd}}{\NA} \alpha^4
\right. \nonumber \\
&& \left. ~~~~
+ \frac{3}{32} \zeta_4 C_A^4 \alpha^2
+ \frac{4}{3} \zeta_3 C_A^2 T_F^2 \Nf^2 \alpha
+ \frac{5}{8} \zeta_5 \frac{d_A^{abcd} d_A^{abcd}}{\NA} \alpha^3
+ \frac{5}{96} \zeta_5 C_A^4 \alpha^3
\right. \nonumber \\
&& \left. ~~~~
+ \frac{9}{128} \zeta_4 C_A^4 \alpha
+ \frac{19}{1536} C_A^4 \alpha^4
+ \frac{33}{32} \zeta_3 \frac{d_A^{abcd} d_A^{abcd}}{\NA} \alpha^2
+ \frac{111}{16} \zeta_3 \frac{d_A^{abcd} d_A^{abcd}}{\NA} \alpha
\right. \nonumber \\
&& \left. ~~~~
+ \frac{271}{1536} C_A^4 \alpha^3
+ \frac{353}{768} \zeta_3 C_A^4 \alpha^2
+ \frac{1039}{384} \zeta_3 C_A^4 \alpha
+ \frac{3283}{2304} C_A^4 \alpha^2
\right. \nonumber \\
&& \left. ~~~~
+ \frac{635749}{62208} C_A^4 \alpha
+ 3 \zeta_4 C_A^2 C_F T_F \Nf \alpha
\right] a^4 ~+~ O(a^5) ~.
\end{eqnarray}
By this we mean that the coefficients of $a$ are in one-to-one correspondence
with the simple poles of $Z_\alpha$. The anomalous dimension of ${\cal O}$ is
\begin{eqnarray}
\gamma_{\cal O}(a,\alpha) &=&
\left[
\frac{4}{3} T_F \Nf
- \frac{35}{12} C_A
+ \frac{1}{4} C_A \alpha
\right] a
\nonumber \\
&&
+ \left[
4 C_F T_F \Nf
- \frac{449}{48} C_A^2
+ \frac{1}{16} C_A^2 \alpha^2
+ \frac{11}{16} C_A^2 \alpha
+ \frac{35}{6} C_A T_F \Nf
\right] a^2
\nonumber \\
&&
+ \left[
12 \zeta_3 C_A C_F T_F \Nf
- \frac{75607}{1728} C_A^3
- \frac{193}{27} C_A T_F^2 \Nf^2
- \frac{44}{9} C_F T_F^2 \Nf^2
+ \frac{3}{8} \zeta_3 C_A^3 \alpha
\right. \nonumber \\
&& \left. ~~~~
+ \frac{3}{128} C_A^3 \alpha^3
+ \frac{9}{32} \zeta_3 C_A^3
+ \frac{101}{256} C_A^3 \alpha^2
+ \frac{167}{64} C_A^3 \alpha
+ \frac{415}{36} C_A C_F T_F \Nf
\right. \nonumber \\
&& \left. ~~~~
+ \frac{5563}{108} C_A^2 T_F \Nf
- 9 \zeta_3 C_A^2 T_F \Nf
- 2 C_F^2 T_F \Nf
- C_A^2 T_F \Nf \alpha
\right] a^3
\nonumber \\
&&
+ \left[
60 \zeta_5 C_A^2 C_F T_F \Nf
- \frac{29939021}{124416} C_A^4
- \frac{175123}{1944} C_A^2 C_F T_F \Nf
- \frac{47665}{1024} \zeta_5 C_A^4
\right. \nonumber \\
&& \left. ~~~~
- \frac{16117}{243} C_A C_F T_F^2 \Nf^2
- \frac{10235}{1536} \zeta_5 C_A^4 \alpha
- \frac{10185}{128} \zeta_5 \frac{d_A^{abcd} d_A^{abcd}}{\NA}
- \frac{3805}{24} \zeta_3 \frac{d_A^{abcd} d_A^{abcd}}{\NA}
\right. \nonumber \\
&& \left. ~~~~
- \frac{2434}{27} C_A^2 T_F^2 \Nf^2
- \frac{1369}{8} \zeta_3 C_A^3 T_F \Nf
- \frac{1352}{27} C_F^2 T_F^2 \Nf^2
- \frac{1232}{243} C_F T_F^3 \Nf^3
\right. \nonumber \\
&& \left. ~~~~
- \frac{1229}{972} C_A^2 T_F^2 \Nf^2 \alpha
- \frac{922}{243} C_A T_F^3 \Nf^3
- \frac{863}{48} C_A^2 C_F T_F \Nf \alpha
- \frac{808}{9} \zeta_3 C_A C_F T_F^2 \Nf^2
\right. \nonumber \\
&& \left. ~~~~
- \frac{778}{243} C_A^3 T_F \Nf \alpha
- \frac{512}{3} \zeta_3 \Nf^2 \frac{d_F^{abcd} d_F^{abcd}}{\NA}
- \frac{512}{9} \Nf \frac{d_F^{abcd} d_A^{abcd}}{\NA}
- \frac{475}{64} \zeta_5 \frac{d_A^{abcd} d_A^{abcd}}{\NA} \alpha
\right. \nonumber \\
&& \left. ~~~~
- \frac{203}{128} \zeta_4 C_A^4 \alpha
- \frac{115}{288} C_A^3 T_F \Nf \alpha^2
- \frac{99}{64} \zeta_4 C_A^4
- \frac{7}{4} \zeta_4 C_A^3 T_F \Nf \alpha
- \frac{5}{48} \zeta_3 C_A^3 T_F \Nf \alpha^2
\right. \nonumber \\
&& \left. ~~~~
- \frac{5}{192} \zeta_3 C_A^4 \alpha^3
- \frac{3}{8} \frac{d_A^{abcd} d_A^{abcd}}{\NA} \alpha
- \frac{3}{32} \frac{d_A^{abcd} d_A^{abcd}}{\NA} \alpha^2
- \frac{1}{1536} \zeta_3 C_A^4 \alpha^4
\right. \nonumber \\
&& \left. ~~~~
+ \frac{1}{16} \zeta_3 \frac{d_A^{abcd} d_A^{abcd}}{\NA} \alpha^3
+ \frac{1}{64} \zeta_3 \frac{d_A^{abcd} d_A^{abcd}}{\NA} \alpha^2
+ \frac{1}{64} \zeta_3 \frac{d_A^{abcd} d_A^{abcd}}{\NA} \alpha^4
+ \frac{3}{64} \zeta_4 C_A^4 \alpha^2
\right. \nonumber \\
&& \left. ~~~~
+ \frac{4}{3} \zeta_3 C_A^2 T_F^2 \Nf^2 \alpha
+ \frac{5}{12} \zeta_5 C_A^4 \alpha^2
+ \frac{5}{16} \zeta_5 \frac{d_A^{abcd} d_A^{abcd}}{\NA} \alpha^3
+ \frac{5}{192} \zeta_5 C_A^4 \alpha^3
\right. \nonumber \\
&& \left. ~~~~
+ \frac{19}{1536} C_A^4 \alpha^4
+ \frac{32}{9} \zeta_3 C_A T_F^3 \Nf^3
+ \frac{46}{9} \zeta_3 C_A^2 T_F^2 \Nf^2
+ \frac{169}{1536} \zeta_3 C_A^4 \alpha^2
\right. \nonumber \\
&& \left. ~~~~
+ \frac{185}{64} \zeta_5 \frac{d_A^{abcd} d_A^{abcd}}{\NA} \alpha^2
+ \frac{314}{9} \zeta_3 C_A C_F^2 T_F \Nf
+ \frac{355}{1536} C_A^4 \alpha^3
+ \frac{449}{32} \zeta_3 \frac{d_A^{abcd} d_A^{abcd}}{\NA} \alpha
\right. \nonumber \\
&& \left. ~~~~
+ \frac{704}{9} \Nf^2 \frac{d_F^{abcd} d_F^{abcd}}{\NA}
+ \frac{704}{9} \zeta_3 C_F^2 T_F^2 \Nf^2
+ \frac{801}{16} \zeta_4 C_A^3 T_F \Nf
+ \frac{1155}{128} \zeta_3 C_A^4 \alpha
\right. \nonumber \\
&& \left. ~~~~
+ \frac{1448}{9} \zeta_3 C_A^2 C_F T_F \Nf
+ \frac{1520}{3} \zeta_3 \Nf \frac{d_F^{abcd} d_A^{abcd}}{\NA}
+ \frac{1939}{288} \frac{d_A^{abcd} d_A^{abcd}}{\NA}
+ \frac{8765}{4608} C_A^4 \alpha^2
\right. \nonumber \\
&& \left. ~~~~
+ \frac{19255}{108} C_A C_F^2 T_F \Nf
+ \frac{163271}{4608} \zeta_3 C_A^4
+ \frac{2184485}{124416} C_A^4 \alpha
+ \frac{2209297}{5184} C_A^3 T_F \Nf
\right. \nonumber \\
&& \left. ~~~~
- 120 \zeta_5 C_A C_F^2 T_F \Nf
- 66 \zeta_4 C_A^2 C_F T_F \Nf
- 46 C_F^3 T_F \Nf
- 18 \zeta_4 C_A^2 T_F^2 \Nf^2
\right. \nonumber \\
&& \left. ~~~~
- 12 \zeta_3 C_A^3 T_F \Nf \alpha
+ 3 \zeta_4 C_A^2 C_F T_F \Nf \alpha
+ 14 \zeta_3 C_A^2 C_F T_F \Nf \alpha
+ 24 \zeta_4 C_A C_F T_F^2 \Nf^2
\right. \nonumber \\
&& \left. ~~~~
+ 55 \zeta_5 C_A^3 T_F \Nf
+ 60 \zeta_5 \Nf \frac{d_F^{abcd} d_A^{abcd}}{\NA}
\right] a^4 ~+~ O(a^5) ~.
\end{eqnarray}

\sect{Five loop $\mMOM$ $\beta$-function.}

In this appendix we record the full five loop $\mMOM$ $\beta$-function for an
arbitrary gauge parameter and colour group. We have
\begin{eqnarray}
\beta^{\mMOMss}(a,\alpha) &=&
\left[
\frac{4}{3} T_F \Nf
- \frac{11}{3} C_A
\right] a^2
\nonumber \\
&&
+ \left[
4 C_F T_F \Nf
- \frac{34}{3} C_A^2
- \frac{10}{3} C_A T_F \Nf \alpha
- \frac{2}{3} C_A T_F \Nf \alpha^2
- \frac{1}{8} C_A^2 \alpha^3
+ \frac{11}{24} C_A^2 \alpha^2
\right. \nonumber \\
&& \left. ~~~~
+ \frac{20}{3} C_A T_F \Nf
+ \frac{65}{12} C_A^2 \alpha
 \right] a^3
\nonumber \\
&&
+ \left[
\frac{64}{3} \zeta_3 C_F T_F^2 \Nf^2
- \frac{9655}{72} C_A^3
- \frac{2971}{576} C_A^3 \alpha^2
- \frac{185}{6} C_A^2 T_F \Nf \alpha
- \frac{184}{9} C_F T_F^2 \Nf^2
\right. \nonumber \\
&& \left. ~~~~
- \frac{176}{3} \zeta_3 C_A C_F T_F \Nf
- \frac{46}{3} C_A T_F^2 \Nf^2
- \frac{32}{3} \zeta_3 C_A T_F^2 \Nf^2
- \frac{25}{16} C_A^3 \alpha^3
- \frac{15}{8} \zeta_3 C_A^3 \alpha
\right. \nonumber \\
&& \left. ~~~~
- \frac{11}{48} \zeta_3 C_A^3 \alpha^2
- \frac{1}{6} \zeta_3 C_A^2 T_F \Nf \alpha^2
- \frac{1}{16} \zeta_3 C_A^3 \alpha^3
+ \frac{1}{12} C_A^2 T_F \Nf \alpha^4
\right. \nonumber \\
&& \left. ~~~~
+ \frac{13}{6} C_A^2 T_F \Nf \alpha^3
+ \frac{43}{192} C_A^3 \alpha^4
+ \frac{119}{36} C_A^2 T_F \Nf \alpha^2
+ \frac{137}{6} \zeta_3 C_A^2 T_F \Nf
\right. \nonumber \\
&& \left. ~~~~
+ \frac{143}{8} \zeta_3 C_A^3
+ \frac{641}{9} C_A C_F T_F \Nf
+ \frac{2009}{18} C_A^2 T_F \Nf
+ \frac{4075}{96} C_A^3 \alpha
\right. \nonumber \\
&& \left. ~~~~
- 20 C_A C_F T_F \Nf \alpha
- 3 C_A C_F T_F \Nf \alpha^2
- 2 C_F^2 T_F \Nf
 \right] a^4
\nonumber \\
&&
+ \left[
128 C_F T_F^3 \Nf^3
- \frac{1381669}{648} C_A^4
- \frac{312001}{432} C_A^3 T_F \Nf \alpha
- \frac{117185}{432} \zeta_3 C_A^4 \alpha
\right. \nonumber \\
&& \left. ~~~~
- \frac{89663}{2304} C_A^4 \alpha^2
- \frac{35965}{72} \zeta_5 C_A^3 T_F \Nf
- \frac{10925}{2304} C_A^4 \alpha^3
- \frac{9428}{9} C_A C_F T_F^2 \Nf^2
\right. \nonumber \\
&& \left. ~~~~
- \frac{2288}{3} \zeta_3 C_A C_F^2 T_F \Nf
- \frac{2145}{4} C_A^2 C_F T_F \Nf \alpha
- \frac{1760}{3} \zeta_5 C_A^2 C_F T_F \Nf
\right. \nonumber \\
&& \left. ~~~~
- \frac{1655}{3} C_A^2 T_F^2 \Nf^2
- \frac{1436}{9} \zeta_3 C_A^2 T_F^2 \Nf^2
- \frac{1409}{12} \zeta_3 C_A^3 T_F \Nf
- \frac{1280}{3} \zeta_5 C_F^2 T_F^2 \Nf^2
\right. \nonumber \\
&& \left. ~~~~
- \frac{823}{576} \zeta_3 C_A^4 \alpha^2
- \frac{749}{36} \zeta_3 C_A^3 T_F \Nf \alpha^2
- \frac{704}{3} \zeta_3 \frac{d_A^{abcd} d_A^{abcd}}{\NA}
- \frac{527}{9} C_A C_F^2 T_F \Nf
\right. \nonumber \\
&& \left. ~~~~
- \frac{512}{3} \zeta_3 \Nf^2 \frac{d_F^{abcd} d_F^{abcd}}{\NA}
- \frac{512}{9} \Nf \frac{d_F^{abcd} d_A^{abcd}}{\NA}
- \frac{269}{768} C_A^4 \alpha^5
- \frac{256}{3} \zeta_3 C_F T_F^3 \Nf^3
\right. \nonumber \\
&& \left. ~~~~
- \frac{205}{12} \zeta_5 C_A^3 T_F \Nf \alpha
- \frac{175}{36} C_A^3 T_F \Nf \alpha^4
- \frac{64}{3} \zeta_3 C_A C_F T_F^2 \Nf^2 \alpha^2
- \frac{64}{27} C_A T_F^3 \Nf^3 \alpha
\right. \nonumber \\
&& \left. ~~~~
- \frac{5}{9} \zeta_5 C_A^3 T_F \Nf \alpha^3
- \frac{5}{16} \zeta_5 C_A^4 \alpha^4
- \frac{1}{3} C_A^3 T_F \Nf \alpha^5
+ \frac{1}{12} \zeta_3 C_A^3 T_F \Nf \alpha^4
\right. \nonumber \\
&& \left. ~~~~
+ \frac{1}{64} \zeta_3 C_A^4 \alpha^5
+ \frac{3}{4} C_A^2 C_F T_F \Nf \alpha^4
+ \frac{19}{9} C_A^2 C_F T_F \Nf \alpha^2
+ \frac{23}{36} \zeta_3 C_A^3 T_F \Nf \alpha^3
\right. \nonumber \\
&& \left. ~~~~
+ \frac{68}{9} \zeta_3 C_A^2 T_F^2 \Nf^2 \alpha
+ \frac{69}{4} C_A^2 C_F T_F \Nf \alpha^3
+ \frac{80}{9} \frac{d_A^{abcd} d_A^{abcd}}{\NA}
+ \frac{80}{9} \zeta_3 C_A^2 T_F^2 \Nf^2 \alpha^2
\right. \nonumber \\
&& \left. ~~~~
+ \frac{112}{9} C_A^2 T_F^2 \Nf^2 \alpha^2
+ \frac{133}{192} \zeta_3 C_A^4 \alpha^4
+ \frac{173}{3} \zeta_3 C_A^2 C_F T_F \Nf \alpha^2
+ \frac{595}{576} \zeta_5 C_A^4 \alpha^3
\right. \nonumber \\
&& \left. ~~~~
+ \frac{184}{9} C_A C_F T_F^2 \Nf^2 \alpha^2
+ \frac{232}{9} C_F^2 T_F^2 \Nf^2
+ \frac{256}{27} \zeta_3 C_A T_F^3 \Nf^3 \alpha
+ \frac{6355}{96} \zeta_5 C_A^4 \alpha
\right. \nonumber \\
&& \left. ~~~~
+ \frac{460}{3} C_A C_F T_F^2 \Nf^2 \alpha
+ \frac{640}{3} \zeta_5 C_A C_F T_F^2 \Nf^2
+ \frac{704}{9} \Nf^2 \frac{d_F^{abcd} d_F^{abcd}}{\NA}
+ \frac{735}{64} \zeta_5 C_A^4 \alpha^2
\right. \nonumber \\
&& \left. ~~~~
+ \frac{896}{27} \zeta_3 C_A T_F^3 \Nf^3
+ \frac{1024}{3} \zeta_3 C_F^2 T_F^2 \Nf^2
+ \frac{1081}{144} C_A^3 T_F \Nf \alpha^3
+ \frac{1280}{9} \zeta_5 C_A^2 T_F^2 \Nf^2
\right. \nonumber \\
&& \left. ~~~~
+ \frac{1291}{72} C_A^3 T_F \Nf \alpha^2
+ \frac{1376}{3} \zeta_3 C_A C_F T_F^2 \Nf^2
+ \frac{1664}{3} \zeta_3 \Nf \frac{d_F^{abcd} d_A^{abcd}}{\NA}
\right. \nonumber \\
&& \left. ~~~~
+ \frac{1792}{81} C_A T_F^3 \Nf^3
+ \frac{2863}{27} C_A^2 T_F^2 \Nf^2 \alpha
+ \frac{3520}{3} \zeta_5 C_A C_F^2 T_F \Nf
+ \frac{5431}{1152} \zeta_3 C_A^4 \alpha^3
\right. \nonumber \\
&& \left. ~~~~
+ \frac{8975}{2304} C_A^4 \alpha^4
+ \frac{60685}{36} C_A^2 C_F T_F \Nf
+ \frac{85855}{288} \zeta_5 C_A^4
+ \frac{111749}{144} C_A^4 \alpha
\right. \nonumber \\
&& \left. ~~~~
+ \frac{229559}{432} \zeta_3 C_A^4
+ \frac{244805}{108} C_A^3 T_F \Nf
- 680 \zeta_3 C_A^2 C_F T_F \Nf
- 160 \zeta_3 C_A C_F T_F^2 \Nf^2 \alpha
\right. \nonumber \\
&& \left. ~~~~
- 46 C_F^3 T_F \Nf
+ 2 C_A C_F^2 T_F \Nf \alpha^2
+ 4 \zeta_3 C_A^3 T_F \Nf \alpha
+ 15 C_A C_F^2 T_F \Nf \alpha
\right. \nonumber \\
&& \left. ~~~~
+ 435 \zeta_3 C_A^2 C_F T_F \Nf \alpha
\right] a^5
\nonumber \\
&&
+ \left[
24640 \zeta_7 C_A^2 C_F^2 T_F \Nf
- \frac{372847895}{6912} \zeta_5 C_A^4 T_F \Nf
- \frac{361398305}{6912} C_A^5
\right. \nonumber \\
&& \left. ~~~~
- \frac{335190709}{221184} \zeta_7 C_A^5 \alpha
- \frac{95542693}{82944} C_A^5 \alpha^2
- \frac{93650917}{13824} \zeta_3 C_A^5 \alpha
\right. \nonumber \\
&& \left. ~~~~
- \frac{85702099}{442368} \zeta_7 C_A^5 \alpha^2
- \frac{41573035}{55296} \zeta_5 C_A^5 \alpha
- \frac{31354477}{1536} \zeta_7 C_A^5
\right. \nonumber \\
&& \left. ~~~~
- \frac{24862345}{36864} \zeta_5 C_A^5 \alpha^2
- \frac{16052591}{864} C_A^4 T_F \Nf \alpha
- \frac{13787695}{1728} \zeta_3 C_A^4 T_F \Nf
\right. \nonumber \\
&& \left. ~~~~
- \frac{9257251}{576} \zeta_3 C_A \frac{d_A^{abcd} d_A^{abcd}}{\NA}
- \frac{7423031}{4608} \zeta_3^2 C_A^5
- \frac{7261237}{432} C_A^3 C_F T_F \Nf \alpha
\right. \nonumber \\
&& \left. ~~~~
- \frac{6892445}{27648} C_A^5 \alpha^3
- \frac{6019289}{216} C_A^3 T_F^2 \Nf^2
- \frac{3214211}{512} \zeta_7 C_A \frac{d_A^{abcd} d_A^{abcd}}{\NA}
\right. \nonumber \\
&& \left. ~~~~
- \frac{2425645}{110592} \zeta_5 C_A^5 \alpha^4
- \frac{2277977}{3456} \zeta_3 C_A^4 T_F \Nf \alpha^2
- \frac{1777405}{48} \zeta_5 C_A^3 C_F T_F \Nf
\right. \nonumber \\
&& \left. ~~~~
- \frac{1606055}{9216} \zeta_5 C_A^5 \alpha^3
- \frac{1352099}{73728} \zeta_7 C_A^5 \alpha^3
- \frac{1238893}{55296} \zeta_3 C_A^5 \alpha^4
\right. \nonumber \\
&& \left. ~~~~
- \frac{1105007}{36} \zeta_3 C_A^2 C_F^2 T_F \Nf
- \frac{829223}{18} C_A^2 C_F T_F^2 \Nf^2
- \frac{789839}{216} C_A^2 C_F^2 T_F \Nf
\right. \nonumber \\
&& \left. ~~~~
- \frac{434315}{288} \zeta_5 T_F \Nf \frac{d_A^{abcd} d_A^{abcd}}{\NA}
- \frac{389095}{1152} \zeta_3 C_A \frac{d_A^{abcd} d_A^{abcd}}{\NA} \alpha
\right. \nonumber \\
&& \left. ~~~~
- \frac{186035}{144} \zeta_5 T_F \Nf \frac{d_A^{abcd} d_A^{abcd}}{\NA} \alpha
- \frac{176425}{48} \zeta_3^2 T_F \Nf \frac{d_A^{abcd} d_A^{abcd}}{\NA}
\right. \nonumber \\
&& \left. ~~~~
- \frac{146863}{128} \zeta_3^2 C_A \frac{d_A^{abcd} d_A^{abcd}}{\NA} \alpha
- \frac{142513}{36} \zeta_3 C_A^3 C_F T_F \Nf
- \frac{51497}{9216} \zeta_3 C_A^5 \alpha^5
\right. \nonumber \\
&& \left. ~~~~
- \frac{100480}{9} \zeta_3 T_F \Nf^2 \frac{d_F^{abcd} d_A^{abcd}}{\NA}
- \frac{74240}{9} \zeta_5 C_A C_F T_F^3 \Nf^3
- \frac{35968}{9} \zeta_3 C_F^2 T_F^3 \Nf^3
\right. \nonumber \\
&& \left. ~~~~
- \frac{62288}{27} C_A \Nf \frac{d_F^{abcd} d_A^{abcd}}{\NA}
- \frac{61427}{48} \zeta_3 C_A^3 C_F T_F \Nf \alpha^2
- \frac{30800}{3} \zeta_7 C_A^3 C_F T_F \Nf
\right. \nonumber \\
&& \left. ~~~~
- \frac{59840}{3} \zeta_3^2 C_A \Nf \frac{d_F^{abcd} d_A^{abcd}}{\NA}
- \frac{56960}{3} \zeta_5 C_A C_F^2 T_F^2 \Nf^2
- \frac{35200}{3} \zeta_5 C_A^2 C_F^2 T_F \Nf \alpha
\right. \nonumber \\
&& \left. ~~~~
- \frac{51968}{9} \zeta_3 C_A \Nf^2 \frac{d_F^{abcd} d_F^{abcd}}{\NA}
- \frac{33860}{9} \zeta_5 C_A \Nf \frac{d_F^{abcd} d_A^{abcd}}{\NA}
- \frac{30080}{3} \zeta_5 C_F^3 T_F^2 \Nf^2
\right. \nonumber \\
&& \left. ~~~~
- \frac{30779}{768} \zeta_3^2 C_A \frac{d_A^{abcd} d_A^{abcd}}{\NA} \alpha^2
- \frac{25888}{9} \zeta_3 C_A C_F T_F^3 \Nf^3
- \frac{19097}{6} C_A C_F^3 T_F \Nf
\right. \nonumber \\
&& \left. ~~~~
- \frac{24064}{27} T_F \Nf^3 \frac{d_F^{abcd} d_F^{abcd}}{\NA}
- \frac{21647}{288} C_A \frac{d_A^{abcd} d_A^{abcd}}{\NA} \alpha
- \frac{19145}{4} \zeta_3 C_A^3 T_F^2 \Nf^2
\right. \nonumber \\
&& \left. ~~~~
- \frac{21472}{9} \zeta_3 C_A \Nf \frac{d_F^{abcd} d_A^{abcd}}{\NA} \alpha
- \frac{21469}{2048} \zeta_7 C_A \frac{d_A^{abcd} d_A^{abcd}}{\NA} \alpha^4
- \frac{12508}{81} C_A^2 T_F^3 \Nf^3 \alpha
\right. \nonumber \\
&& \left. ~~~~
- \frac{21347}{576} \zeta_3^2 C_A^4 T_F \Nf \alpha
- \frac{19865}{108} \zeta_5 C_A^3 T_F^2 \Nf^2 \alpha^2
- \frac{19061}{1152} \zeta_7 T_F \Nf \frac{d_A^{abcd} d_A^{abcd}}{\NA} \alpha^2
\right. \nonumber \\
&& \left. ~~~~
- \frac{17900}{27} \zeta_5 C_A^3 T_F^2 \Nf^2 \alpha
- \frac{8576}{9} C_F T_F^4 \Nf^4
- \frac{15956}{3} \zeta_3 C_A^2 C_F T_F^2 \Nf^2 \alpha
\right. \nonumber \\
&& \left. ~~~~
- \frac{11840}{9} C_A C_F T_F^3 \Nf^3 \alpha
- \frac{10960}{9} \zeta_5 T_F \Nf^2 \frac{d_F^{abcd} d_A^{abcd}}{\NA}
- \frac{10885}{192} \zeta_7 C_A \frac{d_A^{abcd} d_A^{abcd}}{\NA} \alpha^3
\right. \nonumber \\
&& \left. ~~~~
- \frac{10240}{3} \zeta_3 C_A C_F^2 T_F^2 \Nf^2 \alpha
- \frac{9910}{3} \zeta_5 C_A \Nf \frac{d_F^{abcd} d_A^{abcd}}{\NA} \alpha
- \frac{8261}{1152} \zeta_3^2 C_A^4 T_F \Nf \alpha^2
\right. \nonumber \\
&& \left. ~~~~
- \frac{7936}{9} \zeta_3 T_F \Nf^2 \frac{d_F^{abcd} d_A^{abcd}}{\NA} \alpha
- \frac{7931}{54} T_F \Nf \frac{d_A^{abcd} d_A^{abcd}}{\NA}
- \frac{7568}{81} C_A T_F^4 \Nf^4
\right. \nonumber \\
&& \left. ~~~~
- \frac{7040}{9} C_A \Nf^2 \frac{d_F^{abcd} d_F^{abcd}}{\NA} \alpha
- \frac{6560}{3} \zeta_5 C_A^2 C_F T_F^2 \Nf^2 \alpha
- \frac{6400}{27} \zeta_3 C_A T_F^4 \Nf^4
\right. \nonumber \\
&& \left. ~~~~
- \frac{6400}{3} \zeta_5 C_F \Nf \frac{d_F^{abcd} d_A^{abcd}}{\NA}
- \frac{5632}{3} \zeta_3^2 C_A \Nf^2 \frac{d_F^{abcd} d_F^{abcd}}{\NA}
- \frac{6220}{3} \zeta_5 C_A^2 T_F^3 \Nf^3
\right. \nonumber \\
&& \left. ~~~~
- \frac{5120}{3} \zeta_5 T_F \Nf^3 \frac{d_F^{abcd} d_F^{abcd}}{\NA}
- \frac{5120}{3} \zeta_3 C_F \Nf^2 \frac{d_F^{abcd} d_F^{abcd}}{\NA}
- \frac{4925}{6144} \zeta_3^2 C_A^5 \alpha^2
\right. \nonumber \\
&& \left. ~~~~
- \frac{4609}{576} C_A^5 \alpha^5
- \frac{4593}{128} C_A^4 T_F \Nf \alpha^4
- \frac{4400}{3} \zeta_5 C_A^2 C_F^2 T_F \Nf \alpha^2
\right. \nonumber \\
&& \left. ~~~~
- \frac{4160}{3} C_F \Nf^2 \frac{d_F^{abcd} d_F^{abcd}}{\NA}
- \frac{4081}{24576} \zeta_7 C_A^5 \alpha^4
- \frac{3727}{36} C_A^3 T_F^2 \Nf^2 \alpha^3
\right. \nonumber \\
&& \left. ~~~~
- \frac{3416}{27} \zeta_3 C_A^2 T_F^3 \Nf^3 \alpha^2
- \frac{2816}{3} \zeta_3^2 C_A C_F^2 T_F^2 \Nf^2
- \frac{2597}{4} \zeta_7 C_A \Nf \frac{d_F^{abcd} d_A^{abcd}}{\NA} \alpha
\right. \nonumber \\
&& \left. ~~~~
- \frac{2575}{288} C_A \frac{d_A^{abcd} d_A^{abcd}}{\NA} \alpha^2
- \frac{2563}{144} \zeta_3 C_A^4 T_F \Nf \alpha^3
- \frac{2320}{9} C_A C_F^2 T_F^2 \Nf^2 \alpha
\right. \nonumber \\
&& \left. ~~~~
- \frac{2320}{27} \zeta_5 C_A^2 T_F^3 \Nf^3 \alpha
- \frac{2275}{384} \zeta_5 C_A \frac{d_A^{abcd} d_A^{abcd}}{\NA} \alpha^3
- \frac{2264}{3} \zeta_3 C_A C_F^3 T_F \Nf
\right. \nonumber \\
&& \left. ~~~~
- \frac{2051}{48} C_A^3 C_F T_F \Nf \alpha^4
- \frac{2009}{144} \zeta_7 T_F \Nf \frac{d_A^{abcd} d_A^{abcd}}{\NA} \alpha
- \frac{1867}{54} \zeta_3 C_A^3 T_F^2 \Nf^2 \alpha
\right. \nonumber \\
&& \left. ~~~~
- \frac{1688}{3} \zeta_3 C_A \Nf \frac{d_F^{abcd} d_A^{abcd}}{\NA} \alpha^2
- \frac{1640}{27} \zeta_3 C_A^2 T_F^3 \Nf^3 \alpha
- \frac{1412}{3} \zeta_3 C_A^3 C_F T_F \Nf \alpha^3
\right. \nonumber \\
&& \left. ~~~~
- \frac{1280}{3} \zeta_3 C_F \Nf \frac{d_F^{abcd} d_A^{abcd}}{\NA}
- \frac{1280}{3} \zeta_3 C_A C_F^2 T_F^2 \Nf^2 \alpha^2
- \frac{1243}{3} \zeta_3^2 C_A^3 T_F^2 \Nf^2
\right. \nonumber \\
&& \left. ~~~~
- \frac{1024}{3} \zeta_3^2 C_A C_F T_F^3 \Nf^3
- \frac{1000}{9} C_F^2 T_F^3 \Nf^3
- \frac{880}{9} C_A \Nf^2 \frac{d_F^{abcd} d_F^{abcd}}{\NA} \alpha^2
\right. \nonumber \\
&& \left. ~~~~
- \frac{800}{3} \zeta_5 C_A^2 C_F T_F^2 \Nf^2 \alpha^2
- \frac{712}{3} \zeta_3^2 C_A^2 C_F T_F^2 \Nf^2
- \frac{539}{3} \zeta_7 C_A^3 T_F^2 \Nf^2 \alpha
\right. \nonumber \\
&& \left. ~~~~
- \frac{506}{3} C_A^2 C_F T_F^2 \Nf^2 \alpha^3
- \frac{401}{576} \zeta_3 C_A \frac{d_A^{abcd} d_A^{abcd}}{\NA} \alpha^4
- \frac{375}{2048} \zeta_3 C_A^5 \alpha^6
\right. \nonumber \\
&& \left. ~~~~
- \frac{299}{384} \zeta_3 C_A \frac{d_A^{abcd} d_A^{abcd}}{\NA} \alpha^5
- \frac{290}{9} C_A C_F^2 T_F^2 \Nf^2 \alpha^2
- \frac{257}{72} \zeta_3 T_F \Nf \frac{d_A^{abcd} d_A^{abcd}}{\NA} \alpha^2
\right. \nonumber \\
&& \left. ~~~~
- \frac{256}{27} \zeta_3 C_A^2 T_F^3 \Nf^3 \alpha^3
- \frac{160}{3} \zeta_3^2 C_A^2 C_F T_F^2 \Nf^2 \alpha
- \frac{147}{8} \zeta_7 C_A \Nf \frac{d_F^{abcd} d_A^{abcd}}{\NA} \alpha^2
\right. \nonumber \\
&& \left. ~~~~
- \frac{128}{3} \zeta_3^2 T_F \Nf^2 \frac{d_F^{abcd} d_A^{abcd}}{\NA} \alpha
- \frac{119}{144} \zeta_3 T_F \Nf \frac{d_A^{abcd} d_A^{abcd}}{\NA} \alpha^4
- \frac{85}{4} \zeta_3 C_A^3 C_F T_F \Nf \alpha^4
\right. \nonumber \\
&& \left. ~~~~
- \frac{77}{192} \zeta_3 C_A^4 T_F \Nf \alpha^5
- \frac{43}{12} C_A^3 T_F^2 \Nf^2 \alpha^4
- \frac{41}{24} \zeta_3 C_A \frac{d_A^{abcd} d_A^{abcd}}{\NA} \alpha^3
\right. \nonumber \\
&& \left. ~~~~
- \frac{39}{8} C_A^3 C_F T_F \Nf \alpha^5
- \frac{33}{2} C_A^2 C_F^2 T_F \Nf \alpha^3
- \frac{23}{3} C_A^2 C_F T_F^2 \Nf^2 \alpha^4
\right. \nonumber \\
&& \left. ~~~~
- \frac{13}{72} C_A^2 C_F^2 T_F \Nf \alpha^2
- \frac{10}{3} \zeta_5 C_A^3 C_F T_F \Nf \alpha^3
- \frac{8}{3} \zeta_3 C_A^3 T_F^2 \Nf^2 \alpha^4
\right. \nonumber \\
&& \left. ~~~~
- \frac{8}{9} T_F \Nf \frac{d_A^{abcd} d_A^{abcd}}{\NA} \alpha
- \frac{7}{256} \zeta_3 C_A \frac{d_A^{abcd} d_A^{abcd}}{\NA} \alpha^6
- \frac{5}{9} \zeta_5 C_A^3 T_F^2 \Nf^2 \alpha^3
\right. \nonumber \\
&& \left. ~~~~
- \frac{5}{64} \zeta_3^2 C_A \frac{d_A^{abcd} d_A^{abcd}}{\NA} \alpha^3
- \frac{3}{4} C_A^2 C_F^2 T_F \Nf \alpha^4
- \frac{1}{12} \zeta_3^2 C_A^4 T_F \Nf \alpha^3
\right. \nonumber \\
&& \left. ~~~~
- \frac{1}{16} C_A^3 C_F T_F \Nf \alpha^6
- \frac{1}{24} \zeta_3 T_F \Nf \frac{d_A^{abcd} d_A^{abcd}}{\NA} \alpha^5
- \frac{1}{96} \zeta_3 C_A^4 T_F \Nf \alpha^6
\right. \nonumber \\
&& \left. ~~~~
+ \frac{3}{32} C_A \frac{d_A^{abcd} d_A^{abcd}}{\NA} \alpha^4
+ \frac{5}{18} \zeta_5 C_A^4 T_F \Nf \alpha^5
+ \frac{5}{144} \zeta_5 T_F \Nf \frac{d_A^{abcd} d_A^{abcd}}{\NA} \alpha^4
\right. \nonumber \\
&& \left. ~~~~
+ \frac{5}{64} \zeta_5 C_A^5 \alpha^6
+ \frac{5}{192} \zeta_5 C_A \frac{d_A^{abcd} d_A^{abcd}}{\NA} \alpha^5
+ \frac{7}{18} T_F \Nf \frac{d_A^{abcd} d_A^{abcd}}{\NA} \alpha^2
\right. \nonumber \\
&& \left. ~~~~
+ \frac{9}{64} \zeta_3^2 C_A^4 T_F \Nf \alpha^4
+ \frac{16}{9} \zeta_3 C_A^3 T_F^2 \Nf^2 \alpha^3
+ \frac{21}{256} \zeta_3^2 C_A \frac{d_A^{abcd} d_A^{abcd}}{\NA} \alpha^4
\right. \nonumber \\
&& \left. ~~~~
+ \frac{25}{256} \zeta_3^2 C_A^5 \alpha^5
+ \frac{35}{9} \zeta_3^2 C_A^3 T_F^2 \Nf^2 \alpha^2
+ \frac{40}{27} \zeta_5 C_A^2 T_F^3 \Nf^3 \alpha^2
+ \frac{64}{27} C_A^2 T_F^3 \Nf^3 \alpha^3
\right. \nonumber \\
&& \left. ~~~~
+ \frac{67}{96} C_A \frac{d_A^{abcd} d_A^{abcd}}{\NA} \alpha^3
+ \frac{68}{81} C_A^2 T_F^3 \Nf^3 \alpha^2
+ \frac{73}{48} \zeta_3^2 T_F \Nf \frac{d_A^{abcd} d_A^{abcd}}{\NA} \alpha^2
\right. \nonumber \\
&& \left. ~~~~
+ \frac{115}{2} C_A C_F^3 T_F \Nf \alpha^2
+ \frac{136}{9} \zeta_3^2 C_A^3 T_F^2 \Nf^2 \alpha
+ \frac{137}{144} C_A^4 T_F \Nf \alpha^6
+ \frac{147}{64} \zeta_3^2 C_A^5 \alpha^3
\right. \nonumber \\
&& \left. ~~~~
+ \frac{147}{256} \zeta_7 C_A^4 T_F \Nf \alpha^4
+ \frac{320}{3} \zeta_3 C_A C_F T_F^3 \Nf^3 \alpha^2
+ \frac{424}{3} \zeta_3^2 C_A \Nf \frac{d_F^{abcd} d_A^{abcd}}{\NA} \alpha
\right. \nonumber \\
&& \left. ~~~~
+ \frac{440}{3} \zeta_3^2 C_A^3 C_F T_F \Nf \alpha
+ \frac{441}{1024} \zeta_7 C_A^5 \alpha^5
+ \frac{539}{2} \zeta_7 C_A \Nf \frac{d_F^{abcd} d_A^{abcd}}{\NA}
\right. \nonumber \\
&& \left. ~~~~
+ \frac{611}{3} \zeta_3 C_A^2 C_F T_F^2 \Nf^2 \alpha^2
+ \frac{640}{3} \zeta_3 C_A \Nf^2 \frac{d_F^{abcd} d_F^{abcd}}{\NA} \alpha^2
+ \frac{1003}{2048} \zeta_3^2 C_A^5 \alpha^4
\right. \nonumber \\
&& \left. ~~~~
+ \frac{640}{9} C_A \Nf \frac{d_F^{abcd} d_A^{abcd}}{\NA} \alpha^2
+ \frac{1024}{3} \zeta_3^2 C_F^2 T_F^3 \Nf^3
+ \frac{1148}{3} C_F^3 T_F^2 \Nf^2
\right. \nonumber \\
&& \left. ~~~~
+ \frac{1280}{9} \zeta_3^2 C_A^2 T_F^3 \Nf^3
+ \frac{1600}{3} \zeta_5 C_A C_F^2 T_F^2 \Nf^2 \alpha^2
+ \frac{2048}{3} \zeta_3^2 T_F \Nf^3 \frac{d_F^{abcd} d_F^{abcd}}{\NA}
\right. \nonumber \\
&& \left. ~~~~
+ \frac{2176}{3} \zeta_3 C_F^3 T_F^2 \Nf^2
+ \frac{2225}{1536} \zeta_5 C_A^5 \alpha^5
+ \frac{2720}{3} \zeta_5 T_F \Nf^2 \frac{d_F^{abcd} d_A^{abcd}}{\NA} \alpha
\right. \nonumber \\
&& \left. ~~~~
+ \frac{2560}{27} \zeta_5 C_A T_F^4 \Nf^4
+ \frac{2755}{4608} \zeta_5 C_A \frac{d_A^{abcd} d_A^{abcd}}{\NA} \alpha^4
+ \frac{2771}{8} \zeta_3^2 T_F \Nf \frac{d_A^{abcd} d_A^{abcd}}{\NA} \alpha
\right. \nonumber \\
&& \left. ~~~~
+ \frac{3584}{9} \zeta_3 C_F T_F^4 \Nf^4
+ \frac{4157}{6} C_F^4 T_F \Nf
+ \frac{4186}{9} \zeta_3 C_A^3 T_F^2 \Nf^2 \alpha^2
+ \frac{4681}{576} C_A^4 T_F \Nf \alpha^5
\right. \nonumber \\
&& \left. ~~~~
+ \frac{5120}{3} \zeta_3 C_A \Nf^2 \frac{d_F^{abcd} d_F^{abcd}}{\NA} \alpha
+ \frac{5120}{9} C_A \Nf \frac{d_F^{abcd} d_A^{abcd}}{\NA} \alpha
+ \frac{5120}{9} \zeta_5 C_F T_F^4 \Nf^4
\right. \nonumber \\
&& \left. ~~~~
+ \frac{5995}{8} \zeta_5 C_A^3 C_F T_F \Nf \alpha^2
+ \frac{6515}{288} \zeta_5 C_A^4 T_F \Nf \alpha^3
+ \frac{8221}{18432} C_A^5 \alpha^6
\right. \nonumber \\
&& \left. ~~~~
+ \frac{8960}{9} \zeta_3 C_A C_F T_F^3 \Nf^3 \alpha
+ \frac{9679}{18} C_A^2 C_F T_F^2 \Nf^2 \alpha^2
+ \frac{9728}{9} \zeta_3 T_F \Nf^3 \frac{d_F^{abcd} d_F^{abcd}}{\NA}
\right. \nonumber \\
&& \left. ~~~~
+ \frac{10645}{18} C_A^2 C_F^2 T_F \Nf \alpha
+ \frac{10873}{18} C_A^3 C_F T_F \Nf \alpha^2
+ \frac{11200}{3} \zeta_7 C_A^2 C_F T_F^2 \Nf^2
\right. \nonumber \\
&& \left. ~~~~
+ \frac{11449}{12} \zeta_3 C_A^2 C_F^2 T_F \Nf \alpha^2
+ \frac{11629}{216} C_A^3 T_F^2 \Nf^2 \alpha^2
+ \frac{12245}{288} \zeta_5 T_F \Nf \frac{d_A^{abcd} d_A^{abcd}}{\NA} \alpha^2
\right. \nonumber \\
&& \left. ~~~~
+ \frac{12511}{36} C_A^4 T_F \Nf \alpha^3
+ \frac{12800}{3} \zeta_5 C_F \Nf^2 \frac{d_F^{abcd} d_F^{abcd}}{\NA}
+ \frac{12800}{3} \zeta_5 C_F^2 T_F^3 \Nf^3
\right. \nonumber \\
&& \left. ~~~~
+ \frac{12800}{3} \zeta_5 C_A C_F^2 T_F^2 \Nf^2 \alpha
+ \frac{13313}{3456} \zeta_3 C_A^4 T_F \Nf \alpha^4
+ \frac{17123}{48} C_A^3 C_F T_F \Nf \alpha^3
\right. \nonumber \\
&& \left. ~~~~
+ \frac{17455}{3456} \zeta_5 C_A^4 T_F \Nf \alpha^4
+ \frac{18130}{3} \zeta_5 C_A^3 C_F T_F \Nf \alpha
+ \frac{18976}{27} T_F \Nf^2 \frac{d_F^{abcd} d_A^{abcd}}{\NA}
\right. \nonumber \\
&& \left. ~~~~
+ \frac{21760}{3} \zeta_3^2 T_F \Nf^2 \frac{d_F^{abcd} d_A^{abcd}}{\NA}
+ \frac{22895}{3} \zeta_3 C_A^2 C_F^2 T_F \Nf \alpha
+ \frac{70298}{27} C_A C_F^2 T_F^2 \Nf^2
\right. \nonumber \\
&& \left. ~~~~
+ \frac{29440}{9} \zeta_5 C_A \Nf^2 \frac{d_F^{abcd} d_F^{abcd}}{\NA}
+ \frac{29765}{4608} \zeta_5 C_A \frac{d_A^{abcd} d_A^{abcd}}{\NA} \alpha^2
\right. \nonumber \\
&& \left. ~~~~
+ \frac{50075}{72} \zeta_3 T_F \Nf \frac{d_A^{abcd} d_A^{abcd}}{\NA} \alpha
+ \frac{50771}{9} \zeta_3 C_A^2 C_F T_F^2 \Nf^2
+ \frac{58400}{3} \zeta_5 C_A^2 C_F^2 T_F \Nf
\right. \nonumber \\
&& \left. ~~~~
+ \frac{59675}{3} \zeta_5 C_A^3 T_F^2 \Nf^2
+ \frac{72160}{3} \zeta_5 C_A C_F^3 T_F \Nf
+ \frac{72745}{216} C_A \frac{d_A^{abcd} d_A^{abcd}}{\NA}
\right. \nonumber \\
&& \left. ~~~~
+ \frac{104944}{3} \zeta_3 C_A \Nf \frac{d_F^{abcd} d_A^{abcd}}{\NA}
+ \frac{74396}{27} \zeta_3 C_A^2 T_F^3 \Nf^3
+ \frac{83765}{3072} \zeta_3^2 C_A^5 \alpha
\right. \nonumber \\
&& \left. ~~~~
+ \frac{94906}{27} C_A^2 T_F^3 \Nf^3
+ \frac{106477}{2304} \zeta_7 C_A \frac{d_A^{abcd} d_A^{abcd}}{\NA} \alpha
+ \frac{249463}{384} \zeta_3^2 C_A^4 T_F \Nf
\right. \nonumber \\
&& \left. ~~~~
+ \frac{110768}{27} C_A \Nf^2 \frac{d_F^{abcd} d_F^{abcd}}{\NA}
+ \frac{264899}{48} \zeta_3 T_F \Nf \frac{d_A^{abcd} d_A^{abcd}}{\NA}
+ \frac{827533}{3456} \zeta_3 C_A^5 \alpha^3
\right. \nonumber \\
&& \left. ~~~~
+ \frac{269663}{2304} \zeta_3 C_A \frac{d_A^{abcd} d_A^{abcd}}{\NA} \alpha^2
+ \frac{274201}{36} \zeta_3 C_A^3 C_F T_F \Nf \alpha
\right. \nonumber \\
&& \left. ~~~~
+ \frac{284879}{27} C_A^2 C_F T_F^2 \Nf^2 \alpha
+ \frac{292201}{128} \zeta_7 T_F \Nf \frac{d_A^{abcd} d_A^{abcd}}{\NA}
+ \frac{294560}{9} \zeta_5 C_A^2 C_F T_F^2 \Nf^2
\right. \nonumber \\
&& \left. ~~~~
+ \frac{330676}{27} C_A C_F T_F^3 \Nf^3
+ \frac{387247}{4608} \zeta_7 C_A \frac{d_A^{abcd} d_A^{abcd}}{\NA} \alpha^2
+ \frac{477991}{108} C_A^3 T_F^2 \Nf^2 \alpha
\right. \nonumber \\
&& \left. ~~~~
+ \frac{547505}{768} \zeta_5 C_A^4 T_F \Nf \alpha^2
+ \frac{628457}{18432} C_A^5 \alpha^4
+ \frac{1067927}{1152} C_A^4 T_F \Nf \alpha^2
\right. \nonumber \\
&& \left. ~~~~
+ \frac{1299809}{27648} \zeta_7 C_A^4 T_F \Nf \alpha^2
+ \frac{1776235}{432} \zeta_3 C_A^4 T_F \Nf \alpha
+ \frac{1940675}{192} \zeta_3^2 C_A \frac{d_A^{abcd} d_A^{abcd}}{\NA}
\right. \nonumber \\
&& \left. ~~~~
+ \frac{3094355}{384} \zeta_5 C_A \frac{d_A^{abcd} d_A^{abcd}}{\NA}
+ \frac{3105695}{1152} \zeta_5 C_A^4 T_F \Nf \alpha
+ \frac{3115943}{2304} \zeta_3 C_A^5 \alpha^2
\right. \nonumber \\
&& \left. ~~~~
+ \frac{10776505}{2304} \zeta_5 C_A \frac{d_A^{abcd} d_A^{abcd}}{\NA} \alpha
+ \frac{13317479}{384} \zeta_7 C_A^4 T_F \Nf
- 160 C_A C_F T_F^3 \Nf^3 \alpha^2
\right. \nonumber \\
&& \left. ~~~~
+ \frac{20916689}{432} C_A^3 C_F T_F \Nf
+ \frac{82264805}{6912} \zeta_3 C_A^5
+ \frac{346292345}{20736} C_A^5 \alpha
\right. \nonumber \\
&& \left. ~~~~
+ \frac{14551649}{13824} \zeta_7 C_A^4 T_F \Nf \alpha
+ \frac{349180015}{9216} \zeta_5 C_A^5
+ \frac{370399597}{5184} C_A^4 T_F \Nf
\right. \nonumber \\
&& \left. ~~~~
- 24640 \zeta_7 C_A C_F^3 T_F \Nf
- 8960 \zeta_7 C_A C_F^2 T_F^2 \Nf^2
- 145 \zeta_5 C_A \Nf \frac{d_F^{abcd} d_A^{abcd}}{\NA} \alpha^2
\right. \nonumber \\
&& \left. ~~~~
- 9912 \zeta_7 C_A^3 T_F^2 \Nf^2
- 98 \zeta_7 T_F \Nf^2 \frac{d_F^{abcd} d_A^{abcd}}{\NA}
- 8 \zeta_3^2 C_A^2 C_F T_F^2 \Nf^2 \alpha^2
\right. \nonumber \\
&& \left. ~~~~
+ 4 \zeta_3^2 C_A \Nf \frac{d_F^{abcd} d_A^{abcd}}{\NA} \alpha^2
+ 8 \zeta_3 C_A^2 C_F T_F^2 \Nf^2 \alpha^4
+ 22 \zeta_3^2 C_A^3 C_F T_F \Nf \alpha^2
\right. \nonumber \\
&& \left. ~~~~
+ 128 \zeta_3 C_F^4 T_F \Nf
+ 176 \zeta_3 C_A^2 C_F T_F^2 \Nf^2 \alpha^3
+ 196 \zeta_7 T_F \Nf^2 \frac{d_F^{abcd} d_A^{abcd}}{\NA} \alpha
\right. \nonumber \\
&& \left. ~~~~
+ 320 C_F \Nf \frac{d_F^{abcd} d_A^{abcd}}{\NA}
+ 460 C_A C_F^3 T_F \Nf \alpha
+ 3234 \zeta_3^2 C_A^3 C_F T_F \Nf
\right. \nonumber \\
&& \left. ~~~~
+ 8960 \zeta_7 C_F^3 T_F^2 \Nf^2
+ 23440 \zeta_3 C_A C_F^2 T_F^2 \Nf^2
\right] a^6 ~+~ O(a^7) 
\end{eqnarray}
which clearly does not involve $\zeta_4$ or $\zeta_6$. The one loop term is
correctly scheme and gauge parameter independent. We have verified that the 
$\alpha$~$=$~$0$ expression is in complete agreement with the expression 
recorded in the linear covariant gauge determination for the Landau gauge for a
general Lie group \cite{58}. A similar comment applies to the quark mass 
anomalous dimension at five loops.

\end{document}